\documentclass[onecolumn,10pt]{IEEEtran}

\makeatletter

\usepackage[all]{xy}
\usepackage[latin1]{inputenc}
\usepackage{amsfonts,amsmath,amssymb,amsthm}
\usepackage{indentfirst,latexsym,amscd}
\usepackage{amsbsy}
\usepackage{graphicx,multirow,bm}

\newtheorem{lemma}{Lemma}
\newtheorem{thm}{Theorem}
\newtheorem{pro}{Proposition}
\newtheorem{exmp}{Example}

\newtheorem{rem}{Remark}

\title{Construction of Ternary Minimal Linear Codes with Dimension $n+r$ Via Index Codes and The Access Structures of Their Dual Codes}
\author{Haibo Liu, Qunying Liao
\thanks {The work of H.B. Liu is supported by the Sichuan Natural Science Foundation with NO.2024NSFSC0417 and NO.2026NSFSC0138, the work of Q.Y. Liao is supported by the Natural Science Foundation of China with No.12471494.}
\thanks{H.B.Liu, School of Applied Mathematics, Chengdu University of Information Technology,
 \small Chengdu, Sichuan, China (Email:liuhaibo@cuit.edu.cn).}
 \thanks{Q.Y. Liao, School of Mathematical Sciences, Sichuan Normal University,
 \small Chengdu, Sichuan, China (Email:qunyingliao@sicnu.edu.cn)}
}
\date{}

\begin{document}
\baselineskip15pt \maketitle

\begin{abstract}
Recently, minimal linear codes have been extensively studied due to their applications in secret sharing schemes, secure two-party computations, and so on. Constructing minimal linear codes violating the Ashikhmin-Barg condition and then determining their weight distributions have been interesting in coding theory and cryptography. In this paper, we present
a general construction of ternary linear codes of dimension $n+r$, where $n$ and $r$ are positive integers, and derive a
necessary and sufficient condition for the resulting codes to be minimal. We first specialize the construction using index codes
arising from spreads, with three classes of index codes satisfying the minimality condition, we obtain the parameters
and weight enumerators of the corresponding minimal codes, and determine the access structures type of the secret-sharing schemes based
on their dual codes. By the index codes from Krawtchouk polynomials, we then give a second specialization of the construction, with a class of index codes satisfying the minimality condition, we obtain the parameters of the resulting minimal codes, and determine the
access-structure type of the secret-sharing scheme based on each dual code. Several constructed families of minimal codes violate the Ashikhmin-Barg condition.


\end{abstract}
\section{INTRODUCTION}
$ $

Linear codes over finite fields have been extensively studied because of their specific applications in computer and communication systems, data storage devices, and consumer electronics. Minimal linear codes, as a special class of linear codes, have been widely used in secret sharing schemes, secure two-party computation \cite{Carlet-Ding-Yuan}\cite{Yuan-Ding}, and related areas.

Minimal codewords can be used to construct linear-codes-based-access structures in secret sharing schemes. These schemes are protocols that rely on a distribution algorithm and involve a dealer and a group of participants, see \cite{Shamir}. The dealer divides a secret $S$ into different pieces and then distributes them to the participants set $\mathcal{P}$, only authorized subsets of $\mathcal{P}$ (access structure $\Gamma$) can reconstruct the secret by using their respective shares. A set of participants $D$ is called a minimal authorized subset if $D\in\Gamma$ and no proper subset of $D$ belongs to $\Gamma$. In \cite{Massey}, Massey pointed out that the supports of minimal codewords in the dual code can determine the access structure of a secret sharing scheme as well. But to describe the set of minimal codewords of a linear code is quite difficult in general, even in the binary case. To simplify this task, one can try to find linear codes in which very nonzero codeword is minimal, called minimal linear codes. The problem of finding minimal linear codes was first investigated in \cite{Ding-Yuan}. A sufficient condition for a linear code to be minimal is given by the following lemma.
\begin{lemma}\cite{Ashikhmin-Barg}(Ashikhmin-Barg)\label{lem9}
Let $p$ be a prime, then a linear code $\mathcal{C}$ over $\mathbb{F}_p$ is minimal if
\[\frac{w_{\min}}{w_{\max}}>\frac{p-1}{p},\]
where $w_{\min}$ and $w_{\max}$ denote the minimum and maximum nonzero Hamming weights for $\mathcal{C}$, respectively.
\end{lemma}
With the help of Lemma \ref{lem9}, many minimal linear codes were constructed from linear codes with a few weights \cite{Ding1} \cite{Ding2} \cite{Ding-Ding} \cite{Fu-Klove-Luo-Wei} \cite{Tang-Li-Qi-Zhou-Helleseth} \cite{Xiang} \cite{xu2025r}. The sufficient condition in Lemma \ref{lem9} is not usually necessary for a linear code to be minimal. In recent years, searching for minimal linear codes with $\frac{w_{min}}{w_{max}}\leq \frac{p-1}{p}$ has been an interesting research topic. In 2018, Chang and Hyun \cite{Chang-Hyun} made a breakthrough and constructed an infinite family of minimal binary linear codes with $\frac{w_{min}}{w_{max}}\leq \frac{1}{2}$ by the generic construction
\begin{eqnarray}\label{shizi1}
\mathcal{C}_f=\{(uf(\mathbf{x})+\mathbf{v}\cdot \mathbf{x})_{\mathbf{x}\in {\mathbb{F}_p^n}^\ast}:u\in \mathbb{F}_p,\mathbf{v}\in\mathbb{F}_p^n\}.
\end{eqnarray}
Based on the generic construction, a lot of minimal linear codes are obtained, which do not satisfy the Ashikhmin-Barg condition. Ding et al. \cite{Ding-Heng-Zhou} gave a necessary and sufficient condition for a binary linear code to be minimal, and then employed special Boolean functions to obtain three classes of minimal binary linear codes. Heng et al. \cite{Heng-Ding-Zhou} used the characteristic function of a subset in $\mathbb{F}_3^n$ to construct a class of minimal ternary linear codes. Bartoli and Bonini \cite{Bartoli-Bonini} generalized the construction of minimal linear codes in \cite{Heng-Ding-Zhou} from ternary case to be odd characteristic case. Bonini and Borello \cite{Bonini-Borello} presented a family of minimal codes arising from cutting blocking sets. Tao et al. \cite{Tao-Feng-Li} obtained three-weight or four-weight minimal linear codes by using partial difference sets. In order to improve the dimension $\mathcal{C}_f$, Liu and Liao \cite{Liu-Liao1} employed two Boolean functions to obtain a family of minimal binary linear codes with dimension $n+2$. And then, Liu et al. generalized this construction to the ternary case \cite{Liu2026}, by choosing special functions, and obtained a class of ternary minimal linear codes violating the Ashikhmin-Barg condition, and determine their complete weight enumerators. Recently, based on three Boolean functions, Shaikh et al \cite{Shaikh} presented a family of minimal binary linear codes with dimension $n+3$.
Meanwhile, there are other ways to construct minimal linear codes violating the Ashikhmin-Barg condition, see \cite{alfarano2022three} \cite{borello2024geometric}  \cite{Bartoli-Bonini-Gunes} \cite{cardinali2025minimal} \cite{Li-Yue} \cite{Liu-Liao} \cite{Lu-Wu-Cao} \cite{Mesnager} \cite{Mesnager1} \cite{Tang-Qi-Liao-Zhou} \cite{Xu-Qu} \cite{Zhang}.

The minimal codes obtained from generalizations of the generic construction (\ref{shizi1}) that violate the Ashikhmin-Barg condition have dimension $n+r$ with $1\leq r\leq 3$,
which leads to a relatively low code rate, and the access structure type of the secret sharing scheme based on their dual codes has not yet been identified. To improve the code rate, this paper presents a general construction of minimal codes of dimension $n+r$ with $r\leq n$. By two special classes of index codes, we obtain two explicit constructions from this general construction, derive the parameters and weight enumerators of the corresponding minimal codes, and determine the access structure type of the secret sharing scheme based on their dual codes.

The rest of the paper is organized as follows. Section II reviews the definitions and basic results that will be used in the sequel. Section III presents a generic construction for ternary linear codes of dimension $n+r$, and establishes a necessary and sufficient condition for these codes to be minimal. Using an index code based on partial spreads, Section IV presents the first explicit family of these codes, and derives the minimality condition for this family from the perspective of the index code. With three classes of index codes satisfying this condition, Section V obtains the parameters and weight enumerators of the corresponding minimal codes, which violate the Ashikhmin-Barg condition, and determines the access structure types of the secret sharing schemes based on their dual codes. Using an index code based on Krawtchouk polynomials, Section VI provides the second explicit family of these codes, and derives its minimality condition from the index-code perspective. With a class of index codes satisfying this condition, Section VI obtains the parameters of the corresponding minimal code violating the Ashikhmin-Barg condition, and determines the access structure type of the secret sharing scheme based on its dual code. Section VII concludes the paper.

\section{Preliminaries}
$ $

Throughout the whole paper, let $p$ be a prime, and denote by $\mathbb{F}_p$ the finite field with $p$ elements. An $[n,k,d]$ linear code $\mathcal{C}$ over $\mathbb{F}_ p$ is a $k$-dimensional subspace of $\mathbb{F}_p^n$ with minimum (Hamming) distance $d$. Let $G=(\mathbf{g}_0,\mathbf{g}_1,\cdots,\mathbf{g}_{n-1})$ be a generator matrix of $\mathcal{C}$, i.e., the row vectors of $G$ generate the linear subspace $\mathcal{C}$. For any $i=1,\dots,n$, $A_i$ denotes the number of codewords with Hamming weight $i$ in $\mathcal{C}$ of length $n$. The $\emph{weight}$ $\emph{enumerator}$ of $\mathcal{C}$ is defined by
\[1+A_1z+A_2z^2+\cdots+A_nz^n.\]
The $\emph{weight}$ $\emph{distribution}$ $(1,A_1,\dots,A_n)$ is an important research topic in coding theory, since it contains some crucial information as to estimate the error-correction capability and the probability of error-detection and correction with respect to some algorithms \cite{Ding-Ding}. $\mathcal{C}$ is said to be a $t$-weight code if the number of nonzero $A_i$ in the sequence $(A_1,\dots,A_n)$ is equal to $t$.

For a vector $\mathbf{a}=(a_1,\dots,a_n)\in \mathbb{F}_p^n$, the support of $\mathbf{a}$ is defined by
\[\text{Supp}(\mathbf{a})=\{1\leq i\leq n:a_i\neq 0\}.\]
Let $wt(\mathbf{a})$ be the Hamming weight of $\mathbf{a}$, then $wt(\mathbf{a})=|\text{Supp}(\mathbf{a})|$. For $\mathbf{b}\in \mathbb{F}_p^n$, we say that $\mathbf{a}$ covers $\mathbf{b}$ if $\text{Supp}(\mathbf{b})\subseteq \text{Supp}(\mathbf{a})$,
denoted by $\mathbf{b}\preceq \mathbf{a}$. If $\mathbf{b}\preceq \mathbf{a}$, the relation between $wt(\mathbf{a})$ and $wt(\mathbf{b})$ is given below.
\begin{lemma}\cite{Heng-Ding-Zhou}\label{lem2}
For any $\mathbf{a},\mathbf{b}\in \mathbb{F}_p^m$, $\mathbf{b}\preceq \mathbf{a}$ if and only if
\[\sum_{c \in \mathbb{F}_p^\ast} \mathrm{wt}(\mathbf{a}+c \mathbf{b})=(p-1) \mathrm{wt}(\mathbf{a})-\mathrm{wt}(\mathbf{b}).\]
\end{lemma}
A codeword $\mathbf{c}$ in $\mathcal{C}$ is minimal if $\mathbf{c}$ covers only those codewords $u\mathbf{c}$ $(u\in \mathbb{F}_p^\ast)$. $\mathcal{C}$ is said to be minimal if every codeword in $\mathcal{C}$ is minimal.

In the secret sharing scheme based on $\mathcal{C}$, the secret is an element of $\mathbb{F}_p$, which is called the secret space, and $n-1$ participants $P_1, P_2,\cdots, P_{n-1}$ and a dealer are involved. The dealer is a trusted person, to compute the shares with respect to a secret $s$, the dealer chooses randomly a vector $\mathbf{a} = (a_0,a_1,\cdots, a_{k-1})\in\mathbb{F}_p^k$ such that $s = \mathbf{a}\mathbf{g}_0$. There are altogether $p^{k-1}$ such vectors $\mathbf{a}\in\mathbb{F}_p^k$. The dealer then treats $\mathbf{a}$ as an information vector and computes the corresponding codeword $\mathbf{t}=(t_0,t_1,\cdots,t_{n-1})=\mathbf{a}G$, and gives $t_i$ to participant $P_i$ as share for each $i\geq1$. Since $t_0= \mathbf{a}\mathbf{g}_0 = s$, a set of shares $(t_{i_1},t_{i_2},\cdots,t_{i_m})$, $1\leq i_1<\cdots<i_m\leq n-1$ and $1\leq m\leq n-1$, determines the secret if and only if $\mathbf{g}_0$ is a linear combination of $\mathbf{g}_{i_1},\cdots,\mathbf{g}_{i_m}$. This describes the secret sharing scheme based on $\mathcal{C}$. Naturally, we have also the secret sharing scheme based on the dual code $\mathcal{C}^\bot$. The following lemma describes properties of the minimal access sets of the secret sharing scheme based on $\mathcal{C}^\bot$ \cite{Yuan-Ding}.
\begin{lemma}\cite{Yuan-Ding}
Let $\mathcal{C}$ be an $[n,k,d]$ code, and let $G=(\mathbf{g}_0,\mathbf{g}_1,\cdots,\mathbf{g}_{n-1})$ be its generator matrix, where all $g_i$ are nonzero. If each nonzero codeword of $\mathcal{C}$ is minimal, then in the secret sharing scheme based on $\mathcal{C}^\bot$, there are altogether $p^{k-1}$ minimal access sets. In addition, we have the following.

1  If $\mathbf{g}_i$ is a scalar multiple of $\mathbf{g}_0$, $1\leq i\leq n-1$, then participant $P_i$ must be in every minimal access set. Such a participant is called a dictatorial participant.

2  If $\mathbf{g}_i$ is not a scalar multiple of $\mathbf{g}_0$, $1\leq i\leq n-1$, then participant $P_i$ must be in $(p-1)p^{k-2}$ out of $p^{k-1}$ minimal access sets.
\end{lemma}
The relationship between $\mathcal{C}$ and $\mathcal{C}^\bot$ yields a natural classification of the associated access structures. If $d(\mathcal{C}^\bot)=2$, then there exist two distinct columns of a generator matrix of $\mathcal{C}$ that are proportional, namely, $\mathbf{g}_i=u\mathbf{g}_j$ for some $i\neq j$ and $u\in\mathbb{F}_p^\ast$. After a suitable coordinate permutation (if necessary, swapping the $0$-th and $j$-th coordinates), one of these columns can be chosen as the dealer coordinate $\mathbf{g}_0$. The other proportional column, after relabeling, say $\mathbf{g}_i$ with $i\neq 0$, then satisfies $\mathbf{g}_i=u\mathbf{g}_0$. Hence, up to a relabeling of the participants, the secret-sharing scheme based on $\mathcal{C}^\bot$ has a dictatorial participant, and we call the corresponding access structure Type-I. By contrast, if $d(\mathcal{C}^\bot)\geq3$, there do not exist two proportional columns, and we call the corresponding access structure Type-II.

The Krawtchouk polynomial was first introduced by Lloyd in 1929 \cite{Lloyd}. It has been applied in coding theory, cryptography and combinatorics \cite{krawtchouk}. Here we only give a brief introduction to the Krawtchouk polynomial with the essential properties. For more details, see \cite{krawtchouk} \cite{Lloyd}. Let $n$, $h$ be positive integers and $x$ be a variable taking nonnegative values. The \emph{Krawtchouk polynomial }(of degree $k$ with parameters $h\geq2$ and $n$ ) is defined by
\[K_t^h(x,n)=\sum_{j=0}^t(-1)^j(h-1)^{t-j}\binom{x}{j}\binom{n-x}{t-j}.\]
Accordingly, the \emph{Lloyd polynomial } $\Psi_k^h(x,n)$ (of degree $t$ with parameters $h$ and $n$) is given by
\[\Psi_k^h(x,n)=\sum_{t=0}^kK_t^h(x,n).\]
The following Lemma \ref{lem3} will be useful in the sequel.
\begin{lemma}\cite{Heng-Ding-Zhou}\label{lem3}
Let symbols and notations be defined as the above. Suppose that $1\leq x\leq n$, $1\leq k\leq n-1$, $\mathbf{w}\in \mathbb{Z}_p^n$ with Hamming weight $wt(\mathbf{w})=i$. Then the followings hold,

1 $\Psi_k^h(x,n)=K_k^h(x-1,n-1)$;

2 $K_t^h(0,n)=(h-1)^t\binom{n}{t}$;

3 $|\Psi_k^h(x,n)|\leq(h-1)^k\binom{n-1}{k}$;

4 $\sum_{\mathbf{v}\in  \mathbb{Z}_p^n,wt(\mathbf{v})=t}\zeta_p^{\mathbf{w}\cdot\mathbf{v}}=K_t^p(i,n)$,\\
where $\zeta_p=\textrm{e}^{\frac{2\pi\sqrt{-1}}{p}}$ is a primitive $p$-th root of unity, and the inner product $\mathbf{w}\cdot\mathbf{v}$ in $\mathbb{Z}_p^n$ is defined by $$\mathbf{w}\cdot\mathbf{v}=u_1v_1+\cdots+u_nv_n.$$
\end{lemma}
Note that the upper bound for $|\Psi_k^h(x,n)|$ in Lemma \ref{lem3} is tight, since
\[\Psi_k^h(1,n)=K_k^h(0,n-1)=(h-1)^k\binom{n-1}{k}.\]
This paper focuses on the case $h=3$, for convenience, denote $K_t^3(x,n)$ and $\Psi_k^3(x,n)$ to be $K_t(x,n)$ and $\Psi_k(x,n)$ respectively in the rest of the paper. Assume $f(\mathbf{x})$ is a function from $\mathbb{F}_p^n$ to $\mathbb{F}_p$ such that $f(\mathbf{0})=0$ and $f(\mathbf{b})\neq 0$ for at least one $\mathbf{b}\in \mathbb{F}_p^n$. For any $\mathbf{w}\in \mathbb{F}_p^n$, recall that the Walsh transform of $f$ is given by
\[\widehat{f}(\mathbf{w})=\sum_{\mathbf{x}\in\mathbb{F}_p^n}\zeta_p^{f(\mathbf{x})-\mathbf{w}\cdot \mathbf{x}}.\]

\section{A general construction of ternary linear codes with dimension $n+r$ from functions}

In this section, let $f_1,\ldots,f_r$ be functions from $\mathbb{F}_3^n$ to $\mathbb{F}_3$ with $f_i(0)=0$, and define
\[\mathcal F=\text{Span}_{\mathbb{F}_3}\{f_1,\ldots,f_r\},\]
such that the following conditions hold for any $\varphi\in \mathcal{F}$.

$\bullet$(C1) $f_1,\ldots,f_r$ are linearly independent over $\mathbb{F}_3$;

$\bullet$(C2) if $\varphi\in\mathcal{F}$ and $\varphi(\mathbf{x})=\mathbf{v}\cdot \mathbf{x}$ for every $\mathbf{x}\in\mathbb{F}_3^n$, then $\varphi=0$ and $\mathbf{v}=\mathbf{0}$.

We now define a linear code $\mathcal{C}_{\mathcal{F}}$ by
\begin{eqnarray}\label{shizi2}
\mathcal{C}_{\mathcal{F}}=\left\{\mathbf{c}_{\mathbf{u}}(\mathbf{v})=(\sum_{i=1}^ru_if_i(\mathbf{x})+\mathbf{v} \cdot \mathbf{x})_{\mathbf{x} \in {\mathbb{F}_3^n}^\ast}: \mathbf{u}=(u_1,\ldots,u_r) \in \mathbb{F}_3^r, \mathbf{v} \in \mathbb{F}_3^n\right\}.
\end{eqnarray}
Note that the codes in \cite{Heng-Ding-Zhou} and \cite{Liu2026} are a subcode of $\mathcal{C}_{\mathcal{F}}$, and the dimension and weight distribution of $\mathcal{C}_{\mathcal{F}}$ are established in the following theorem.

\begin{thm}\label{thm1}
Let the notations be defined as above, then $\mathcal{C}_{\mathcal{F}}$ of (\ref{shizi2}) has length $3^n-1$ and dimension $n+r$. Moreover, the weight distribution of $\mathcal{C}_{\mathcal{F}}$ is given by the following multiset union,
\begin{eqnarray*}
& \left\{\left\{2\left(3^{n-1}-\frac{{\textrm Re}(\widehat{\sum_{i=1}^ru_if_i(\mathbf{x})}(-\mathbf{v}))}{3}\right): \mathbf{u} \in {\mathbb{F}_3^r}^\ast, \mathbf{v} \in \mathbb{F}_3^n, \right\}\right\} \bigcup \left\{\left\{3^n-3^{n-1}: \mathbf{u}=\mathbf{0}, \mathbf{v} \in{\mathbb{F}_3^n}^\ast\right\}\right\} \cup\{\{0\}\} .
\end{eqnarray*}
\end{thm}
\begin{proof} For a codeword $\mathbf{c}_{\mathbf{u}}(\mathbf{v})=(\sum_{i=1}^ru_if_i(\mathbf{x})+\mathbf{v} \cdot \mathbf{x})_{\mathbf{x} \in {\mathbb{F}_3^n}^\ast}\in \mathcal{C}_{\mathcal{F}}$, the Hamming weight of $\mathbf{c}_{\mathbf{u}}(\mathbf{v})$ can be calculated as

\begin{eqnarray*}
wt(\mathbf{c}_{\mathbf{u}}(\mathbf{v})) &=&\sharp\left\{\mathbf{x} \in {\mathbb{F}_3^n}^\ast: \sum_{i=1}^ru_if_i(\mathbf{x})+\mathbf{v} \cdot \mathbf{x} \neq 0 \right\} \\
&=&\left(3^n-1\right)-\frac{1}{3} \sum_{y \in \mathbb{F}_3} \sum_{x \in {\mathbb{F}_3^n}^\ast} \zeta_3^{y(\sum_{i=1}^ru_if_i(\mathbf{x})+\mathbf{v} \cdot \mathbf{x})} \\
&=&\left(3^n-1\right)+1-3^{n-1}-\frac{1}{3} \sum_{y \in {\mathbb{F}_3}^\ast} \sum_{x \in \mathbb{F}_3^n} \zeta_3^{y(\sum_{i=1}^ru_if_i(\mathbf{x})+\mathbf{v} \cdot \mathbf{x})}\\
&=&3^n-3^{n-1}-\frac{1}{3}\left(\sum_{x \in \mathbb{F}_3^n} \zeta_3^{\sum_{i=1}^ru_if_i(\mathbf{x})+\mathbf{v} \cdot \mathbf{x}}+\sum_{x \in \mathbb{F}_3^n} \zeta_3^{-\sum_{i=1}^ru_if_i(\mathbf{x})-\mathbf{v} \cdot \mathbf{x}}\right) \\
&=&3^n-3^{n-1}-\frac{2}{3} \textrm{Re}\left(\sum_{x \in \mathbb{F}_3^n} \zeta_3^{\sum_{i=1}^ru_if_i(\mathbf{x})+\mathbf{v} \cdot\mathbf{ x}}\right) .
\end{eqnarray*}
The weight distribution of $\mathcal{C}_{\mathcal{F}}$ follows from the above discussions.

Next, we determine the dimension of $\mathcal{C}_{\mathcal{F}}$, consider the map
\[\Phi:\mathbb{F}_3^r\times\mathbb{F}_3^n\longrightarrow \mathcal{C}_{\mathcal{F}}, \qquad (\mathbf{u},\mathbf{v})\longmapsto \mathbf{c}_{\mathbf{u}}(\mathbf{v}),\]
the map is linear and surjective by the definition of $\mathcal{C}_{\mathcal{F}}$. It remains to determine its kernel. Suppose $\Phi(\mathbf{u},\mathbf{v})=\mathbf{0}$, then
\[\sum_{i=1}^ru_i f_i(\mathbf{x})+\mathbf{v}\cdot \mathbf{x}=\mathbf{0} \quad\text{for $\mathbf{x}\neq\mathbf{0}$}. \]
The same identity is true at $\mathbf{x}=\mathbf{0}$, since every $f_i(\mathbf{0})=0$ and the linear form $\mathbf{v}\cdot \mathbf{x}$ vanishes
there. Thus, on all $\mathbf{x}$ of $\mathbb{F}_3^n$,
\[\sum_{i=1}^ru_i f_i(\mathbf{x})=(-\mathbf{v})\cdot\mathbf{x},\]
Assumption C2 gives $\mathbf{v}=\mathbf{0}$ and $\sum_i^ru_if_i=0$, and then Assumption C1 gives
$u_1=\cdots=u_r=0$. Hence $\ker\Phi=\{\mathbf{0}\}$, so $\Phi$ is bijective and $\dim(\mathcal{C}_{\mathcal{F}})=n+r$.

This completes the proof of Theorem \ref{thm1}.
\end{proof}

To determine the type of access structure based on the dual code of $\mathcal{C}_{\mathcal{F}}$, we need to know $d(\mathcal{C}_{\mathcal{F}}^\bot)$. And the lower bound of $d(\mathcal{C}_{\mathcal{F}}^\bot)$ is given by the following proposition.
\begin{pro}\label{Pro1}
For the code $\mathcal{C}_{\mathcal{F}}$ of (\ref{shizi2}), let $\mathbf{F}(\mathbf{x})=(f_1(\mathbf{x}),\ldots,f_r(\mathbf{x}))^T$, if $\mathbf{F}(2\mathbf{x})=2\mathbf{F}(\mathbf{x})$ for some $\mathbf{x}\in{\mathbb{F}_3^n}^\ast$, then $d\bigl(\mathcal{C}_{\mathcal{F}}^\perp\bigr)=2$, otherwise $d\bigl(\mathcal{C}_{\mathcal{F}}^\perp\bigr)\geq3$.
\end{pro}
\begin{proof} It is easy to check that the generator matrix of $\mathcal{C}_{\mathcal{F}}$ in (\ref{shizi2}) is
\[G=(\mathbf{g}_{\mathbf{x}})_{\mathbf{x} \in {\mathbb{F}_3^n}^\ast}=\binom{\mathbf{F}(\mathbf{x})}{\mathbf{x}}_{\mathbf{x} \in {\mathbb{F}_3^n}^\ast}.\]
Since every column of $G$ lower block is $\mathbf{x}$, then every column of $G$ is nonzero, so $d\bigl(\mathcal{C}_{\mathcal{F}}^\perp\bigr)$ is at least $2$. $d\bigl(\mathcal{C}_{\mathcal{F}}^\perp\bigr)=2$ if and only if two distinct columns of $G$ are proportional, i.e.,
$\mathbf{g}_\mathbf{y}=\lambda \mathbf{g}_\mathbf{x}$ for $\lambda\in\mathbb{F}_3^\ast$. Comparison of the two columns lower blocks gives $\mathbf{y}=\lambda \mathbf{x}$, and the columns are distinct, so $\lambda=2$ and $\mathbf{y}=2\mathbf{x}$. Comparison of the upper blocks then gives $F(2\mathbf{x})=2F(\mathbf{x})$. If the $\mathbf{g}_{\mathbf{2x}}=2\mathbf{g}_\mathbf{x}$ never occurs, then $d\bigl(\mathcal{C}_{\mathcal{F}}^\perp\bigr)\geq3$.

This completes the proof of Proposition \ref{Pro1}.
\end{proof}

A natural question is when the code $\mathcal{C}_{\mathcal{F}}$ is minimal, we will give a necessary and sufficient condition for $\mathcal{C}_{\mathcal{F}}$ to be minimal in term of the Walsh transform of $\mathcal{F}$.
\begin{lemma}\label{lem1}
Let $\mathbf{a}=(a_1,\ldots,a_n),\mathbf{b}=(b_1,\ldots,b_n)\in\mathbb{F}_3^n$, then
\begin{eqnarray}\label{shizi4}
wt(\mathbf{b}+\mathbf{a})+wt(\mathbf{b}-\mathbf{a})-2wt(\mathbf{b})+wt(\mathbf{a})=3\,|\text{Supp}(\mathbf{a})\setminus \text{Supp}(\mathbf{b})|,
\end{eqnarray}
in particular
\[\text{Supp}(\mathbf{a})\subseteq\text{Supp}(\mathbf{b})\Longleftrightarrow
wt(\mathbf{b}+\mathbf{a})+wt(\mathbf{b}-\mathbf{a})=2wt(\mathbf{b})-wt(\mathbf{a}).\]
\end{lemma}
\begin{proof}
For a statement $P$, denote its indicator by $\mathbf1_P$. For $1\leq j\leq n$, the contribution of coordinate $j$ to
the left side of (\ref{shizi4}) is
\[d_j=\mathbf1_{b_j+a_j\neq0}+\mathbf1_{b_j-a_j\neq0}-2\mathbf1_{b_j\neq0}+\mathbf1_{a_j\neq0}.\]
There are three cases to consider. If $a_j=0$, then $d_j=2\mathbf1_{b_j\neq0}-2\mathbf1_{b_j\neq0}=0$.
If $a_j\neq0$ and $b_j\neq0$, then $a_j=b_j$ or $a_j=-b_j$, exactly one of $b_j+a_j$ and $b_j-a_j$ is nonzero, and
$d_j=1-2+1=0$. Finally, if $a_j\neq0$ and $b_j=0$, so
$d_j=1+1+1=3$. Therefore
\[d_j=3\mathbf1_{\{a_j\neq0,\ b_j=0\}},\]
and (\ref{shizi4}) holds. When $\text{Supp}(\mathbf{a})\subseteq \text{Supp}(\mathbf{b})$, the set on the right side of (\ref{shizi4}) is empty.

This completes the proof of Lemma \ref{lem1}.
\end{proof}
Note here that when $p=3$, Lemma \ref{lem2} is a special case of Lemma \ref{lem1}. For two pairs $(\mathbf{u}_1,\mathbf{v}_1)$ and $(\mathbf{u}_2,\mathbf{v}_2)\in \mathbb{F}_3^{r+n}$, define the \emph{Walsh containment defect} by
\begin{equation}
\begin{split}
\Delta((\mathbf{u}_1,\mathbf{v}_1),(\mathbf{u}_2,\mathbf{v}_2))=3^n+2Re_{\mathbf{u}_2}(-\mathbf{v}_2)-Re_{\mathbf{u}_1}(-\mathbf{v}_1)
-Re_{\mathbf{u}_2+\mathbf{u}_1}(-\mathbf{v}_2-\mathbf{v}_1)-Re_{\mathbf{u}_2-\mathbf{u}_1}(-\mathbf{v}_2+\mathbf{v}_1),
\end{split}
\end{equation}
where
\[Re_{\mathbf{u}}(-\mathbf{v})=Re(\widehat{\sum_{i=1}^ru_if_i}(-\mathbf{v})).\]
Now, based on \emph{Walsh containment defect}, we present a necessary and sufficient condition that $\mathcal{C}_{\mathcal{F}}$ is minimal.
\begin{thm}\label{thm2}
Let $\mathcal{C}_{\mathcal{F}}$ be the ternary code of Theorem \ref{thm1}, then $\mathcal{C}_{\mathcal{F}}$ is minimal if and only if
\[\Delta((\mathbf{u}_1,\mathbf{v}_1),(\mathbf{u}_2,\mathbf{v}_2))>0.\]
for any two ordered linearly independent vectors $(\mathbf{u}_1,\mathbf{v}_1)$, $(\mathbf{u}_2,\mathbf{v}_2)\in \mathbb{F}_3^{r+n}$ over $\mathbb{F}_3$.
\end{thm}
\begin{proof}
For convenience, let $\mathbf{c}_1=\mathbf{c}_{\mathbf{u}_1}(\mathbf{v}_1)$ and $\mathbf{c}_2=\mathbf{c}_{\mathbf{u}_2}(\mathbf{v}_2)$, then \[\mathbf{c}_2\pm\mathbf{c}_1=\mathbf{c}_{\mathbf{u}_2\pm\mathbf{u}_1}(\mathbf{v}_2\pm \mathbf{v}_1).\]
According to Lemma \ref{lem1}, we have
\[3\,|\text{Supp}(\mathbf{c}_1)\setminus \text{Supp}(\mathbf{c}_2)|=wt(\mathbf{c}_2+\mathbf{c}_1)+wt(\mathbf{c}_2-\mathbf{c}_1)-2wt(\mathbf{c}_2)+wt(\mathbf{c}_1).\]
By Theorem \ref{thm1} $wt(\mathbf{c}_{\mathbf{u}}(\mathbf{v}))=\frac{2}{3}(3^n-Re_{\mathbf{u}}(\mathbf{-v}))$, then
\begin{eqnarray*}
3\,|\text{Supp}(\mathbf{c}_1)\setminus \text{Supp}(\mathbf{c}_2)|&=&\frac{2}{3}(3^n+2Re_{\mathbf{u}_2}(-\mathbf{v}_2)-Re_{\mathbf{u}_1}(-\mathbf{v}_1)\\
&-&Re_{\mathbf{u}_2+\mathbf{u}_1}(-\mathbf{v}_2-\mathbf{v}_1)-Re_{\mathbf{u}_2-\mathbf{u}_1}(-\mathbf{v}_2+\mathbf{v}_1))\\
&=&\frac{2}{3}\Delta((\mathbf{u}_1,\mathbf{v}_1),(\mathbf{u}_2,\mathbf{v}_2)).
\end{eqnarray*}
It is clear that $|\text{Supp}(\mathbf{c}_1)\setminus \text{Supp}(\mathbf{c}_2)|\geq 0$, and $(\mathbf{u}_1,\mathbf{v}_1)$, $(\mathbf{u}_2,\mathbf{v}_2)$ are linear independent over $\mathbb{F}_3$, then
\[|\text{Supp}(\mathbf{c}_1)\setminus \text{Supp}(\mathbf{c}_2)|=\frac{2}{9}\Delta((\mathbf{u}_1,\mathbf{v}_1),(\mathbf{u}_2,\mathbf{v}_2))=0\] if and only if $\mathbf{c}_1\preceq \mathbf{c}_2$. For $\mathbf{c}_2\preceq \mathbf{c}_1$ case, one can similarly obtain
\[|\text{Supp}(\mathbf{c}_2)\setminus \text{Supp}(\mathbf{c}_1)|=\frac{2}{9}\Delta((\mathbf{u}_2,\mathbf{v}_2),(\mathbf{u}_1,\mathbf{v}_1))=0.\]

This completes the proof of Theorem \ref{thm2}.
\end{proof}

\section{A Class of Ternary Minimal Linear Codes from Partial Spread-Based Index Codes}

In this section, using the general construction (\ref{shizi2}), we shall present a class of minimal ternary codes from a class of specific functions based on partial spreads. We always assume that $n$ is a positive even integer in this section, denote $\frac{n}{2}=m$, $M=3^m$ and $Q=M^2$. Let $\mathcal{L}=\{V_1,\ldots,V_l\}$ be partial spread of order $l$ in $\mathbb{F}_3^{2m}$ such that every $V_j$ has dimension $m$ and $V_j\cap V_k=\{\mathbf{0}\}$ for $j\neq k$. For the explicit family, we use a full spread in this section, for which $l=M+1$. Let $\mathcal{D}\subseteq \mathbb{F}_3^{M+1}$ be an $[M+1,r]$ index code, and let $\mathbf{d}_1, \ldots,\mathbf{d}_r$ be a basis of $\mathcal{D}$ over $\mathbb{F}_3$. For $\mathbf{d}_i=(d_{i,1},d_{i,2},\ldots,d_{i,M+1})$, $1\leq i\leq r$, define the function $f_i:\mathbb{F}_3^n\longrightarrow \mathbb{F}_3$ by
\begin{eqnarray}\label{shizi5}
f_i(\mathbf{x})=\left\{\begin{array}{ll}d_{i,j}, &\text{if}\quad \mathbf{x}\in {V_j}^\ast;\\
 0, &\text{otherwise}, \end{array}\right.
\end{eqnarray}
where ${V_j}^\ast=V_j\setminus \{\mathbf{0}\}$. To indicate the relationship between $f_i$ and $\mathbf{d}_i$, we denote $f_i$ by $f_{\mathbf{d}_i}$. Based on $f_{\mathbf{d}_i}(1\leq i\leq r)$, the general construction (\ref{shizi2}) is used to obtain the ternary linear code $\mathcal{C}_{\mathcal{F}}$. And the weight enumerator of $\mathcal{C}_{\mathcal{F}}$ is given as follows according to the weight distribution of $\mathcal{D}$.

\begin{thm}\label{thm3}
Let notations defined as above, then the code $\mathcal{C}_{\mathcal{F}}$ in (\ref{shizi2}) is a $[3^n-1,n+r]$ ternary linear code, and the weight enumerator of $\mathcal{C}_{\mathcal{F}}$ is given as follows
\[W_{\mathcal{C}_\mathcal{F}}(z)=1+(M^2-1)z^{\frac{2}{3}Q}+\sum_{\mathbf{d}\in \mathcal{D}\setminus \{\mathbf{0}\}}\left(z^{wt(\mathbf{d})(M-1)}+(M+1-wt(\mathbf{d}))(M-1)z^{\frac{2}{3}Q-wt(\mathbf{d})}
+wt(\mathbf{d})(M-1)z^{\frac{2}{3}Q+M-wt(\mathbf{d})}\right),\]
where $wt(\mathbf{d})$ is hamming weight of $\mathbf{d}$.
\end{thm}
\begin{proof}
We first show the code $\mathcal{F}$ satisfies the assumption C1-C2, then the dimension of $\mathcal{C}_{\mathcal{F}}$ is $n+r$ by Theorem \ref{thm1}. According to (\ref{shizi5}), consider the map
\[\Psi:\mathcal{D}\longrightarrow \mathcal{F}, \mathbf{d}\longmapsto f_\mathbf{d}, \]
where $\mathcal{F}=\text{Span}_{\mathbb{F}_3}\{f_1,\ldots,f_r\}$, obviously, $\Psi$ is a bijection, and $\mathcal{F}$ satisfies the assumption C1. For any $\mathbf{d}\neq \mathbf{0}$, choose $j\in \text{Supp}(\mathbf{d})$, let $\mathbf{x}\in {V_j}^\ast$, then $2\mathbf{x}\in {V_j}^\ast$, by (\ref{shizi5}), we have
\[f_{\mathbf{d}}(2\mathbf{v})=f_{\mathbf{d}}(\mathbf{v})\neq2f_\mathbf{d}(\mathbf{v}),\]
thus $f_{\mathbf{d}}$ is not linear, and $\mathcal{F}$ satisfies the assumption C2.

Next, we calculate the weight enumerator of $\mathcal{C}_{\mathcal{F}}$. By (\ref{shizi5}), for nonzero vector $\mathbf{b}=(b_1,\ldots,b_{M+1})\in\mathcal{D}$, there exists $f_\mathbf{b}\in\mathcal{F}$ such that
\begin{eqnarray*}
f_\mathbf{b}(\mathbf{x})=\left\{\begin{array}{ll}b_j, &\text{if}\quad \mathbf{x}\in {V_j}^\ast;\\
 0, &\text{otherwise}. \end{array}\right.
\end{eqnarray*}
For $\mathbf{v}\in \mathbb{F}_3^n$, in order to obtain $wt(\mathbf{c}_{\mathbf{b}}(\mathbf{v}))$, where $\mathbf{c}_{\mathbf{b}}(\mathbf{v})=(f_\mathbf{b}(\mathbf{x})+\mathbf{x}\cdot\mathbf{v})_{\mathbf{x}\in{\mathbb{F}_3^n}^\ast}$, by Theorem \ref{thm1}, we need to compute $\widehat{f_\mathbf{b}(-\mathbf{v})}$. For convenience of computation, write $J_\mathbf{b}=\text{Supp}(\mathbf{b})$, $u_\mathbf{b}=|J_\mathbf{b}|$, and
\[S(\mathbf{b})=\sum_{j\in J_\mathbf{b}}(1-\zeta_3^{b_j}).\]
Then
$$
\begin{aligned}
\widehat{{f}_\mathbf{b}}(-\mathbf{v})&=\sum_{\mathbf{x} \in \mathbb{F}_3^n} \zeta_3^{f_{\mathbf{b}}(\mathbf{x})+\mathbf{v} \cdot \mathbf{x}}
=\sum_{j\in J_\mathbf{b}}\sum_{\mathbf{x}\in {V_j}^\ast}\zeta_3^{b_j+\mathbf{x}\cdot\mathbf{v}}
+\sum_{\mathbf{x}\in\mathbb{F}_3^n\setminus\bigcup_{j\in J_\mathbf{b}}{V_j}^\ast}\zeta_3^{\mathbf{x}\cdot\mathbf{v}}\\
&=\sum_{j\in J_\mathbf{b}}(\zeta_3^{b_j}-1)\sum_{\mathbf{x}\in {V_j}^\ast}\zeta_3^{\mathbf{x}\cdot\mathbf{v}}+\sum_{\mathbf{x}\in\mathbb{F}_3^n}\zeta_3^{\mathbf{x}\cdot\mathbf{v}}.
\end{aligned}
$$
For $\mathbf{v} \in \mathbb{F}_3^n$, if $\mathbf{v}= \mathbf{0}$, then
\[\widehat{f_{\mathbf{b}}}(\mathbf{0})=Q-(M-1)S(\mathbf{b}),\]
for $b_j\in\mathbb{F}_3^\ast$, $Re(\zeta_3^{b_j}-1)=-\frac{3}{2}$, thus
\[Re(\widehat{f_{\mathbf{b}}}(\mathbf{0}))=Q-\frac{3}{2}u_\mathbf{b}(M-1).\]
If $\mathbf{v}\neq \mathbf{0}$, the full-spread property gives $V_j\cap V_k=\{\mathbf{0}\}$ and
$V_j+V_k=\mathbb{F}_3^{2m}$ for $j\neq k$, therefore
$V_j^\perp\cap V_k^\perp=(V_j+V_k)^\perp=\{\mathbf{0}\}$.
So a nonzero vector $\mathbf{v}$ is orthogonal to at most one spread, thus we have
\begin{eqnarray*}
\widehat{{f}_\mathbf{b}}(-\mathbf{v})&=& \begin{cases}\sum_{j\in J_\mathbf{b}\setminus\{k\}}(\zeta_3^{b_j}-1)\sum_{\mathbf{x}\in {V_j}^\ast}\zeta_3^{\mathbf{x}\cdot\mathbf{v}}+(\zeta_3^{b_k}-1)\sum_{\mathbf{x}\in {V_k}^\ast}\zeta_3^{\mathbf{x}\cdot\mathbf{v}}, & \text { if there exists $k\in J_\mathbf{b}$ such that $\mathbf{v}\in{V_k}^\perp$}, \\
\sum_{j\in J_\mathbf{b}}(\zeta_3^{b_j}-1)\sum_{\mathbf{x}\in {V_j}^\ast}\zeta_3^{\mathbf{x}\cdot\mathbf{v}}, & \text {otherwise},\end{cases}\\
&=&\begin{cases}-\sum_{j\in J_\mathbf{b}}(\zeta_3^{b_j}-1)+M(\zeta_3^{b_k}-1), & \text { if there exists $k\in J_\mathbf{b}$ such that $\mathbf{v}\in{V_k}^\perp$}, \\
-\sum_{j\in J_\mathbf{b}}(\zeta_3^{b_j}-1), & \text {otherwise}.\end{cases}
\end{eqnarray*}
Thus
\begin{eqnarray*}
Re(\widehat{{f}_\mathbf{b}}(-\mathbf{v}))&=&\begin{cases}\frac{3}{2}(u_{\mathbf{b}}-M), & \text { if there exists $k\in J_\mathbf{b}$ such that $\mathbf{v}\in{V_k}^\perp$}, \\
\frac{3}{2}u_{\mathbf{b}}, & \text {otherwise}.\end{cases}
\end{eqnarray*}
The weight of the code $\mathbf{c}_{f_\mathbf{b}}(\mathbf{v})$ follows from Theorem \ref{thm1}.
Note that the number of $\mathbf{v}$ such that there exists $k\in J_\mathbf{b}$ with $\mathbf{v}\in{V_k}^\perp$ is $u_{\mathbf{b}}(M-1)$, the number of the remaining $\mathbf{v}$ is $(M+1-u_{\mathbf{b}})(M-1)$.

If $\mathbf{b}=\mathbf{0}\in\mathcal{D}$, then $\mathbf{c}_{\mathbf{0}}(\mathbf{v})=(\mathbf{v}\cdot\mathbf{x})_{\mathbf{x}\in{\mathbb{F}_3^n}^\ast}$, and $wt(\mathbf{c}_{\mathbf{0}}(\mathbf{v}))=\frac{2}{3}Q$ for $\mathbf{v}\neq \mathbf{0}$. The number of $\mathbf{v}\neq \mathbf{0}$ is $Q-1$.

This completes the proof of Theorem \ref{thm3}.
\end{proof}

Next, based on Theorem \ref{thm2}, we will show the condition for $\mathcal{C}_{\mathcal{F}}$ to be a minimal code in Theorem \ref{thm3} from the perspective of $\mathcal{D}$.
\begin{thm}\label{thm4}
Assume $n\geq 4$, and let the notations be defined as above, if $\mathcal{D}$ satisfies the following two conditions simultaneously,

(A1) the index code $\mathcal{D}$ is a minimal ternary code,

(A2) every nonzero $\mathbf{b}\in\mathcal{D}$ satisfies $2\leq wt(\mathbf{b})\leq M-2$,

then the linear code $\mathcal{C}_{\mathcal{F}}$ of Theorem \ref{thm3} is minimal.
\end{thm}
\begin{proof}
Let $\mathbf{c}_1$ and $\mathbf{c}_2\in\mathcal{C}_{\mathcal{F}}$ be arbitrary nonzero codewords, for convenience, denote
\[\mathbf{c}_i=(f_{\mathbf{b}_i}(\mathbf{x})+\mathbf{v}_i\cdot \mathbf{x})_{\mathbf{x} \in {\mathbb{F}_3^n}^\ast}\
\text{and}\ J_{\mathbf{b}_i}=\text{Supp}(\mathbf{b}_i),\]
where $\mathbf{b}_i=(b_{i,1},\ldots,b_{i,M+1})\in \mathcal{D}$ and $i=1,2$.
Assume that $\text{Supp}(\mathbf{c}_1)\subseteq \text{Supp}(\mathbf{c}_2)$, under the conditions A1-A2, we will show that either $\mathbf{c}_1$ and $\mathbf{c}_2$ are proportional or the assumption does not hold. According to the relation between $\mathbf{v}_1$ and $\mathbf{v}_2$, we shall discuss this by considering the following four cases.

{\bf Case 1}. $\mathbf{v}_1=\mathbf{v}_2=\mathbf{0}$. The support of $\mathbf{c}_i$ is the disjoint union of the sets
${V_j}^\ast$ with $j\in J_{\mathbf{b}_i}$. Therefore
\[ \text{Supp}(\mathbf{c}_1)\subseteq\text{Supp}(\mathbf{c}_2) \quad\Longleftrightarrow\quad J_{\mathbf{b}_1}\subseteq J_{\mathbf{b}_2}.\]
Since $\mathbf{c}_1$ and $\mathbf{c}_2$ are nonzero, then the index words $\mathbf{b}_1$ and $\mathbf{b}_2$ are nonzero, and the minimality of $\mathcal{D}$ in the condition A1 gives
$\mathbf{b}_1=\lambda\mathbf{b}_2$ for some $\lambda\in{\mathbb{F}_3}^\ast$. Hence $\mathbf{c}_1=\lambda \mathbf{c}_2$.

{\bf Case 2}. Exactly one of $\mathbf{v}_i(i=1,2)$ is zero. We first assume that $\mathbf{v}_1=\mathbf{0}$ and $\mathbf{v}_2\neq\mathbf{0}$.  Fix $j\in J_{\mathbf{b}_1}$, the codeword $\mathbf{c}_1$ is
nonzero at every point $\mathbf{x}$ with $\mathbf{x}\in{V_j}^\ast$, so $\text{Supp}(\mathbf{c}_1)\subseteq \text{Supp}(\mathbf{c}_2)$ requires $\mathbf{c}_2$ to be
nonzero at every such point. If $\mathbf{v}_2\notin V_j^\perp$, we consider nonzero solutions of the equation
\[b_{2,j}+\mathbf{v}_2\cdot \mathbf{x}=0\]
on $V_j$. The restriction $\mathbf{x}\mapsto \mathbf{v}_2\cdot \mathbf{x}$ is a nonzero linear function on the $m$-dimensional space $V_j$,
it is therefore surjective, and every fiber has $3^{m-1}$ elements. If $b_{2,j}\neq0$, the equation has $3^{m-1}$ nonzero solutions; if
$b_{2,j}=0$,the equation has $3^{m-1}-1>0$ nonzero solutions because $n\geq4$. Thus the equation has nonzero solutions,
At each such solution $\mathbf{c}_2(\mathbf{x})=0$, a contradiction. Hence $\mathbf{v}_2\in V_j^\perp$ for every
$j\in J_{\mathbf{b}_1}$. But a nonzero vector belongs to at most one $V_j^\perp$, while
$|J_{\mathbf{b}_1}|\geq2$ by the condition A2.  This case is impossible.

Conversely, suppose that $\mathbf{v}_1\neq\mathbf{0}$ and $\mathbf{v}_2=\mathbf{0}$. For every $j\notin J_{\mathbf{b}_2}$, the codeword
$\mathbf{c}_2$ vanishes on all of ${V_j}^\ast$. $\text{Supp}(\mathbf{c}_1)\subseteq \text{Supp}(\mathbf{c}_2)$ forces $\mathbf{c}_1$ to vanish
identically there. Thus
\[b_{1,j}+\mathbf{v}_1\cdot \mathbf{x}=0\quad\text{and} \quad b_{1,j}+\mathbf{v}_1\cdot 2\mathbf{x}=0 \quad(\mathbf{x}\in {V_j}^\ast, j\notin J_{\mathbf{b}_2}).\]
Comparing this identity at $\mathbf{x}$ and $2\mathbf{x}$ shows that $\mathbf{v}_1\cdot \mathbf{x}=0$ and then
$b_{1,j}=0$, equivalently, $\mathbf{v}_1\in V_j^\perp$. Condition A2 gives
\[|\{j:j\notin J_{\mathbf{b}_2}\}|=(M+1)-wt(\mathbf{b}_2)\geq3,\]
but a nonzero $\mathbf{v}_1$ is in at most one $V_j^\perp$. This is again impossible.

{\bf Case 3}. $\mathbf{v}_1,\mathbf{v}_2$ are nonzero and linearly dependent. There is a $\lambda\in\mathbb{F}_3^\ast$ such that $\mathbf{v}_2=\lambda \mathbf{v}_1$, and
$\lambda \mathbf{c}_1=(f_{\lambda\mathbf{b}_1}(\mathbf{x})+\lambda\mathbf{v}_1\cdot \mathbf{x})_{\mathbf{x} \in {\mathbb{F}_3^n}^\ast}=(f_{\lambda\mathbf{b}_1}(\mathbf{x})+\mathbf{v}_2\cdot \mathbf{x})_{\mathbf{x} \in {\mathbb{F}_3^n}^\ast}$ preserves its support, then we may assume $\mathbf{v}_1=\mathbf{v}_2=\mathbf{v}$. There is at most one index $j$ with $\mathbf{v}\in V_j^\perp$.
For $j_1\in J_{\mathbf{b}_2}$ with $j_1\neq j$, the equation $b_{2,j_1}+\mathbf{v}\cdot \mathbf{x}=0$ has nonzero solutions in $V_{j_1}$ by the same argument as in Case 2.
At such an $\mathbf{x}$, $\mathbf{c}_2(\mathbf{x})=0$, and $\text{Supp}(\mathbf{c}_1)\subseteq \text{Supp}(\mathbf{c}_2)$ forces $\lambda\mathbf{c}_1(\mathbf{x})=\lambda b_{1,j_1}+\mathbf{v}\cdot \mathbf{x}=0$. Subtracting the two equations gives
$\lambda b_{1,j_1}=b_{2,j_1}$. For the other $j_2\notin J_{\mathbf{b}_2}$, if $\mathbf{v}\notin V_{j_2}^\perp$, choose a nonzero $\mathbf{x}\in V_{j_2}$ such that $\mathbf{v}\cdot \mathbf{x}=0$, we obtain $\lambda b_{1,j_2}=b_{2,j_2}=0$ from $\text{Supp}(\mathbf{c}_1)\subseteq \text{Supp}(\mathbf{c}_2)$. If $\mathbf{v}\in V_{j_2}^\perp$, then $\lambda b_{1,j_2}=b_{2,j_2}=0$ by $\text{Supp}(\mathbf{c}_1)\subseteq \text{Supp}(\mathbf{c}_2)$. Consequently
\[wt(\lambda\mathbf{b}_1-\mathbf{b}_2)\leq1.\]
If the difference is nonzero, this would contradict with the fact that the weight of nonzero codewords in $\mathcal{D}$ is at least $2$ by
the condition A2. Hence $\lambda\mathbf{b}_1=\mathbf{b}_2$, and the original codewords $\mathbf{c}_1$ and $\mathbf{c}_2$
are proportional.

{\bf Case 4}. $\mathbf{v}_1,\mathbf{v}_2$ are linearly independent. Then $\mathbf{v}_1,\mathbf{v}_2,\mathbf{v}_2+\mathbf{v}_1$, and $\mathbf{v}_2-\mathbf{v}_1$ are all nonzero. For a nonzero $\mathbf{d}\in \mathcal{D}$ of weight
$u$, from the proof of Theorem \ref{thm3}, we know $Re(f_{\mathbf{d}}(\mathbf{v}))$ is either $\frac{3}{2}u$ or $\frac{3}{2}(u-M)$ at a nonzero $\mathbf{v}$. By the condition A2 $2\leq u\leq M-2$, the values satisfy
\[|Re(f_{\mathbf{d}}(\mathbf{v}))|\leq\frac32(M-2)\qquad(\mathbf{v}\neq\mathbf{0}).\]
The same bound holds for $\mathbf{d}=\mathbf{0}$, because $Re(f_{\mathbf{0}}(\mathbf{v}))=0$ for $\mathbf{v}\neq\mathbf{0}$. If $\text{Supp}(\mathbf{c}_1)\subseteq \text{Supp}(\mathbf{c}_2)$, Theorem \ref{thm2} would make the Walsh containment defect zero, thus
\begin{eqnarray*}
 |Q|&=&|Re_{f_{\mathbf{b}_2+\mathbf{b}_1}}(-\mathbf{v}_2-\mathbf{v}_1)
     +Re_{f_{\mathbf{b}_2-\mathbf{b}_1}}(-\mathbf{v}_2+\mathbf{v}_1)
 -2Re_{f_{\mathbf{b}_2}}(-\mathbf{v}_2)+Re_{f_{\mathbf{b}_1}}(-\mathbf{v}_1)|\\
 &\leq&|Re_{f_{\mathbf{b}_2+\mathbf{b}_1}}(-\mathbf{v}_2-\mathbf{v}_1)|
     +|Re_{f_{\mathbf{b}_2-\mathbf{b}_1}}(-\mathbf{v}_2+\mathbf{v}_1)|
 +2|Re_{f_{\mathbf{b}_2}}(-\mathbf{v}_2)|+|Re_{f_{\mathbf{b}_1}}(-\mathbf{v}_1)|\\
 &\leq&(1+1+2+1)\frac{3}{2}(M-2)=\frac{15}{2}(M-2).
\end{eqnarray*}
Since $n\geq4$, we have $M\geq9$, and
\[Q-\frac{15}{2}(M-2)=M^2-\frac{15}{2}(M-2)=\frac{1}{2}(2M^2-15M+30)>0.\]
This is a contradiction.

From the above discussion, for any two nonzero codewords $\mathbf{c}_1$ and $\mathbf{c}_2$, the assumption $\text{Supp}(\mathbf{c}_1)\subseteq\text{Supp}(\mathbf{c}_2)$ forces that
either $\mathbf{c}_1$ and $\mathbf{c}_2$ are proportional or the assumption does not hold, thus $\mathcal{C}_\mathcal{F}$ is minimal.

This completes the proof of Theorem \ref{thm4}
\end{proof}
To determine the access structure of secret sharing schemes based on $\mathcal{C}_{\mathcal{F}}^\bot$, we need to consider the minimum distance of $\mathcal{C}_{\mathcal{F}}^\bot$, the minimum distance of $\mathcal{C}_{\mathcal{F}}^\bot$ is determined by the generator matrix of $\mathcal{D}$.
\begin{pro}\label{pro2}
Assume that $n\geq4$, let $A$ be an $r\times(M+1)$ generator matrix of the index code $\mathcal{D}$, and let $\mathbf{a}_j$ denote
$j$th column of $A$ with $1\leq j\leq M+1$, then the minimum distance of $\mathcal{C}_\mathcal{F}^\perp$ is
\begin{equation*}
 d\bigl(\mathcal{C}_\mathcal{F}^\perp\bigr)
 =\begin{cases}
 2,&\mathbf{a}_j=0\text{ for some $j$,}\\
 3,&\mathbf{a}_j\neq0\text{ for every $j$}.
 \end{cases}
\end{equation*}
\end{pro}
\begin{proof}
Treat the chosen rows of $A$ as a basis of $\mathcal{D}$, for $1\leq j\leq M+1$, by (\ref{shizi5}), the column of the generator matrix of
$\mathcal{C}_{\mathcal{F}}$ indexed by $\mathbf{x}\in {V_j}^\ast$ is
\[g_\mathbf{x}=\binom{\mathbf{a}_j}{\mathbf{x}}\in\mathbb{F}_3^{r+2m}.\]
No such column is zero because of $\mathbf{x}\neq\mathbf{0}$, so $d\bigl({\mathcal{C}_\mathcal{F}}^\perp\bigr)\geq 2$.

Suppose two distinct columns $g_\mathbf{x}$ and $g_\mathbf{y}$ are proportional. Their last $2m$ entries give
$\mathbf{y}=\lambda \mathbf{x}$ for some $\lambda\in{\mathbb{F}_3}^\ast$, scalar multiplication preserves a spread component,
so $\mathbf{x},\mathbf{y}\in V_j$ for the same $j$. Since $\mathbf{y}$ and $\mathbf{x}$ are distinct, then $\lambda=2$ and
$\mathbf{y}=2\mathbf{x}$. The first $r$ entries then require $\mathbf{a}_j=2\mathbf{a}_j$, equivalently $\mathbf{a}_j=\mathbf{0}$. Conversely,
if $\mathbf{a}_j=\mathbf{0}$, then $F(\mathbf{x})=F(2\mathbf{x})=0$
for every $\mathbf{x}\in {V_j}^\ast$, where $\mathbf{F}(\mathbf{x})=(f_1(\mathbf{x}),\ldots,f_r(\mathbf{x}))^T$. Thus, by Proposition
\ref{Pro1}, the first case of Proposition \ref{pro2} holds.

Now assume that every $\mathbf{a}_j$ is nonzero. The preceding argument shows that no two columns are
proportional, so $d\bigl({\mathcal{C}_\mathcal{F}}^\perp\bigr)\geq 3$. Choose any $1\leq j\leq M+1$ and two linearly
independent vectors $\mathbf{x},\mathbf{y}\in V_j$, which is possible because $m\geq2$, and put $\mathbf{z}=-\mathbf{x}-\mathbf{y}\in V_j$.
The vectors $\mathbf{x},\mathbf{y},\mathbf{z}$ are nonzero and distinct, and
\[g_\mathbf{x}+g_\mathbf{y}+g_\mathbf{z}=\binom{3\mathbf{a}_j}{\mathbf{x}+\mathbf{y}+\mathbf{z}}=\mathbf{0}.\]
Hence $d\bigl({\mathcal{C}_\mathcal{F}}^\perp\bigr)=3$.

This completes the proof of Proposition \ref{pro2}.
\end{proof}
By Proposition \ref{pro2}, we can determine the type of access structure of the secret-sharing scheme based on ${\mathcal{C}_\mathcal{F}}^\perp$.
\begin{rem}
The construction in Theorem \ref{thm3} is general, as long as one can obtain an index code $\mathcal{D}$ satisfying Theorem \ref{thm4}, one can use Theorem \ref{thm3} to obtain an $(n+r)$-dimensional minimal code $\mathcal{C}_\mathcal{F}$. Furthermore, if the weight distribution of $\mathcal{D}$ is available, then the weight distribution of $\mathcal{C}_\mathcal{F}$ can also be obtained by Theorem \ref{thm3}.
\end{rem}

\section{Three Infinite and Explicit Families of Minimal Codes from Theorem \ref{thm3}}

In this section, following Theorem \ref{thm4}, we select three classes of specific index code $\mathcal{D}$ satisfying the required conditions, and use Theorem \ref{thm3} to obtain an $(r+n)$-dimensional minimal code $\mathcal{C}_\mathcal{F}$, then give its exact weight enumerator.

The construction in Theorem \ref{thm3} requires $\mathcal{D}$ to have length $M+1$. For a minimal code of length less than $M+1$, zeros can be appended as coordinates to extend its length to $M+1$. Namely, for an $[n,k,d]$ minimal code $\mathcal{C}$ with $n<M+1$, one can obtain an $[M+1,k,d]$ minimal code $\overline{\mathcal{C}}$ by
\begin{equation}\label{shizi3}
\overline{\mathcal{C}}=\{\overline{\mathbf{c}}=(\mathbf{c},\mathbf{0}_{M+1-n})|\mathbf{c}\in \mathcal{C}\},
\end{equation}
where $\mathbf{0}_{M+1-n}$ is zero vector of length $M+1-n$. From (\ref{shizi3}), it can be seen that (\ref{shizi3}) only changes the code length of $\mathcal{C}$ and does not change the other parameters of the code $\mathcal{C}$. If we take $\mathcal{D}=\overline{\mathcal{C}}$, then by Proposition \ref{pro2}, the access structure of the secret-sharing scheme generated by $\mathcal{C}_{\mathcal{D}}^\bot$ is Type-I.

If we want to obtain a Type-II of access structure based on $\mathcal{C}_{\mathcal{D}}^\bot$, we need to ensure that $\mathcal{C}_{\mathcal{F}}^\bot\geq 3$. Therefore, we cannot use the method of extending the code length in (\ref{shizi3}). Instead, we introduce another way to extend the code length $\mathcal{C}$ without adding zero column vectors. Let
\[G_n=( \mathbf{g}_1,\ldots,\mathbf{ g}_{n})\]
be a generator matrix of a ternary minimal $[n,r]$ code
$\mathcal{C}$. Assume that $\operatorname{rank}(G_n)=r$ and
$\mathbf{g}_i\neq\mathbf{0}$ for every $i$. For arbitrary positive
integers $\mu_1,\ldots,\mu_{n}$, define
\begin{equation}\label{shizi6}
 G_{\boldsymbol\mu}=\left(
 \underbrace{\mathbf{g}_1,\ldots, \mathbf{g}_1}_{\mu_1},
 \ldots,
 \underbrace{ \mathbf{g}_{n},\ldots,\mathbf{g}_{n}}_{\mu_{n}}
 \right),
\end{equation}
where ${\boldsymbol\mu}=(\mu_1,\cdots,\mu_{n})$. Then $G_{\boldsymbol\mu}$ generates a ternary minimal code
$\mathcal{C}_{\boldsymbol\mu}$ of length
$N=\sum_{i=1}^{n}\mu_i$ and dimension $r$. Furthermore, $G_{\boldsymbol\mu}$ has no zero columns. By (\ref{shizi6}), we know that (\ref{shizi6}) changes the code length and weight distribution of $\mathcal{C}$. Next, we explain the relationship of weight between $\mathcal{C}$ and $\mathcal{C}_{\boldsymbol\mu}$. Suppose that the target length of $\mathcal{C}_{\boldsymbol\mu}$ is $M+1=3^m+1$ and $n\leq M$, then
\[M+1=an+s,\qquad 0\leq s<n,\]
choose a set $S$ of $s$ column indices and put
\[\mu_i=\begin{cases}a+1,&i\in S,\\a,&i\notin S.\end{cases}\]
For $\mathbf{u}\in\mathbb{F}_3^r$, define
\[t_{\mathbf{u}}=\#\{i\in S: \mathbf{u}\cdot\mathbf{g}_i\neq0\},\]
then we have
\[wt(\mathbf{u}G_{\boldsymbol\mu})
 =awt(\mathbf{ u}G_n)+t_{\mathbf{u}}.\]
As long as $w_{\min}\geq 2$ and $w_{\max}\leq n-3$, where $w_{\min}$ and $w_{\max}$
are the minimum and maximum nonzero weights of $\mathcal{C}$, $\mathcal{C}_{\boldsymbol\mu}$ satisfies condition (A2) of Theorem \ref{thm4}. In the following, we will choose a specific $\mathcal{C}$ to obtain $\mathcal{D}$ satisfying Theorem \ref{thm4} through the two methods (\ref{shizi3}) and (\ref{shizi6}), and determine the parameters and weight enumerator of $\mathcal{C}_{\mathcal{F}}$.

\subsection{$\mathcal{D}$ is an infinite family derived from simplex codes}

The simplex code is a particularly special class of minimal codes. In this subsection, we first select the simplex code and extend it using (\ref{shizi3}) to obtain the index code $\mathcal{D}$.

For a positive integer $r$, the $[3^r-1,r,2\cdot3^{r-1}]$ simplex code $\mathcal{C}$ is defined by
\[\mathcal{C}=\{\mathbf{c}_{\mathbf{v}}=(\mathbf{x}\cdot \mathbf{v})_{\mathbf{x}\in {\mathbb{F}_3^r}^\ast}:\mathbf{v}\in\mathbb{F}_3^r\},\]
and it is minimal and all of its nonzero codewords have weight $u=2\cdot3^{r-1}$. The generator matrix of $\mathcal{C}$ is
\[G_r=( \mathbf{x})_{ \mathbf{x}\in\mathbb{F}_3^r\setminus\{\mathbf{0}\}}.\]
Thus, every nonzero vector of $\mathbb{F}_3^r$ occurs once as a column, and the matrix has no zero columns.
Based on (\ref{shizi3}), one can obtain an $[M+1,r, 2\cdot3^{r-1}]$ one weight minimal code $\mathcal{D}$. Then, we have the following theorem by Theorem \ref{thm3}.
\begin{thm}\label{thm5}
Suppose that $m\geq 2$ and $2\leq r<m$, with the notations defined as above. Let $\mathcal{D}$ be the index code of $\mathcal{F}$, then $\mathcal{C}_\mathcal{F}$ of Theorem \ref{thm3} is a $[3^{2m}-1,2m+r,2\cdot3^{r-1}(3^m-1)]$ minimal code violating the Ashikhmin-Barg condition, and its weight enumerator is
\begin{eqnarray*}
W_{\mathcal{C}_{\mathcal{F}}}(z)&=&1+(3^{2m}-1)z^{2\cdot3^{2m-1}}+ (3^r-1)z^{2\cdot3^{r-1}(3^m-1)}+2\cdot3^{r-1}(3^r-1)(3^{m}-1)z^{2\cdot3^{2m-1}+3^m-2\cdot3^{r-1}}\\
&+&(3^r-1)(3^m-2\cdot3^{r-1}+1)(3^m-1)z^{2\cdot3^{2m-1}-2\cdot3^{r-1}}.
\end{eqnarray*}
Furthermore, the parameters for $\mathcal{C}_\mathcal{F}^\perp$ is $[3^{2m}-1,3^{2m}-1-2m-r,2]$, and the access structure of the secret-sharing scheme constructed from $\mathcal{C}_\mathcal{F}^\perp$ belongs to Type-I.
\end{thm}
\begin{proof}
The index code $\mathcal{D}$ is a one weight minimal code with the parameters $[M+1,r, 2\cdot3^{r-1}]$, since $m\geq 2$, then
\[2\leq 2\cdot 3^{r-1}<2\cdot3^{m-1}=2\frac{M}{3}\leq M-2.\]
The hypotheses of Theorem \ref{thm4} are therefore
satisfied, which proves $\mathcal{C}_\mathcal{F}$ is minimal. For nonzero codeword $\mathbf{d}\in \mathcal{D}$, $wt(\mathbf{d})=2\cdot3^{r-1}$, the number of $\mathbf{d}$ is $3^r-1$, thus the explicit weight weight enumerator of $\mathcal{C}_\mathcal{F}$ follows from Theorem \ref{thm3}. For convenience, put $u=2\cdot3^{r-1}$, from the weight enumerator of $\mathcal{C}_\mathcal{F}$, we know
\[w_{\min}=u(M-1),\qquad w_{\max}=\frac{2M^2}{3}+M-u.\]
Then
\[\frac{w_{\min}}{w_{\max}}=\frac{u(M-1)}{\frac{2M^2}{3}+M-u},\]
the function $\frac{w_{\min}}{w_{\max}}$ is increasing for $u>0$, $u< \frac{M}{3}$ gives
\[\frac{w_{\min}}{w_{\max}}\leq\frac{(M/3)(M-1)}{2M^2/3+M-M/3}=\frac{M-1}{2(M+1)}<\frac12<\frac23.\]
Thus $\mathcal{C}_\mathcal{F}$ violates the Ashikhmin-Barg condition. Since the code length of $\mathcal{C}$ is $3^r-1<M+1$, (\ref{shizi3}) guarantees there exists at least one zero column in the generator matrix of $\mathcal{D}$, Proposition \ref{pro2} gives $d\bigl(\mathcal{C}_\mathcal{F}^\perp\bigr)=2$, and the access structure of the secret-sharing scheme based on $\mathcal{C}_\mathcal{F}^\perp$ is a Type-I access structure.

This completes the proof of Theorem \ref{thm5}.
\end{proof}
The condition $r<m$ is imposed to guarantee $\frac{w_{\min}}{w_{\max}}<\frac{1}{2}$, which ensures that $\mathcal{C}_\mathcal{F}$ violates the Ashikhmin-Barg condition. However, $r=m$ is attainable in specific examples.
Based on Magma's program, the following example is presented, which is accordant with Theorem \ref{thm5}.
\begin{exmp}
Let $m=3$ and $r=2$, then $\mathcal{D}$ in Theorem \ref{thm5} is a $[28,2,6]$ minimal code with the weight enumerator
 \[W_{\mathcal{D}}(z)=1+8z^6.\]
The code $\mathcal{C}_{\mathcal{F}}$ in Theorem \ref{thm5} is minimal with parameters $[728,8,156]$, and the weight enumerator of $\mathcal{C}_{\mathcal{F}}$ is
\[W_{\mathcal{C}_{\mathcal{F}}}(z)=1+8z^{156}+4576z^{480}+728z^{486}+1248z^{507}.\]
Clearly, $\frac{w_{min}}{w_{max}}=\frac{156}{507}<\frac{2}{3}$. The parameters of the code $\mathcal{C}_{\mathcal{F}}^{\bot}$ in Theorem \ref{thm5} is $[728,720,2]$.
\end{exmp}
Next, we use (\ref{shizi6}) to extend $\mathcal{C}$ to obtain $\mathcal{D}$, such that $d(\mathcal{C}_{\mathcal{F}}^\bot)\geq 3$. Before presenting the conclusion, some preparations are still needed. For every pair of integers $2\leq r\leq m$, write
\[ m=er+t,\qquad e\geq1,\qquad 0\leq t<r,\]
and put
\[ M=3^m,\qquad a=\frac{M-3^t}{3^r-1},\qquad u=2a3^{r-1}.\]
$a$ is a positive integer, because $3^m-3^t$ is divisible by $3^r-1$,
moreover,
\[ M+1=a(3^r-1)+(3^t-1)+2.\]
Fix a $t$-dimensional subspace $U\subseteq\mathbb{F}_3^r$. If $t=0$, put $U=\{\mathbf{0}\}$ and choose any $\mathbf{g}\in\mathbb{F}_3^r\setminus\{\mathbf{0}\}$. If $t\geq1$, choose $\mathbf{g}\in U\setminus\{\mathbf{0}\}$, define
\begin{eqnarray*}
\begin{aligned}
G_{m,r}=&
 \left(
 (\underbrace{ \mathbf{x},\ldots, \mathbf{x}}_{a}
 )_{ \mathbf{x}\in\mathbb{F}_3^r\setminus\{\mathbf{0}\}}
 \mathbin{\|}( \mathbf{x})_{\mathbf{ x}\in U\setminus\{\mathbf{0}\}}
 \mathbin{\|}( \mathbf{g},2\mathbf{g}),\right),
\end{aligned}
\end{eqnarray*}
where $\mathbin{\|}$ denotes column concatenation, the middle block is empty when $t=0$.
This is an $r\times(M+1)$ matrix of rank $r$ with no zero columns. By (\ref{shizi6}), it generates a ternary
minimal code, denoted by $\mathcal{D}_{m,r}=\{\mathbf{d}G_{m,r}|\mathbf{d}\in\mathbb{F}_3^r\}$.

Subsequently, we calculate the weight enumerator of $\mathcal{D}_{m,r}$ according to whether $t=0$ or not. Every nonzero word has a contribution $2a3^{r-1}$ from the first block. If $t=0$, the last two coordinates are both zero or both nonzero, hence $\mathcal{D}_{m,r}$ has parameters $[M+1,r,2a3^{r-1}]$ and the weight enumerator
\[W_{\mathcal{D}_{m,r}}(z)=1+(3^{r-1}-1)z^{2a3^{r-1}}+2\cdot3^{r-1}z^{2a3^{r-1}+2}.\]
If $1\leq t<r$, for a nonzero $\mathbf{d}\in\mathbb{F}_3^r$, $(\mathbf{d}\cdot \mathbf{x})_{\mathbf{ x}\in U\setminus\{\mathbf{0}\}}$ vanishes identically on $U$ gives an extra weight zero, and the number of $\mathbf{d}$ is $3^{r-t}-1$. The remaining $3^{r}-3^{r-t}$ vectors $\mathbf{d}$ give $wt((\mathbf{d}\cdot \mathbf{x})_{\mathbf{ x}\in U\setminus\{\mathbf{0}\}})=2\cdot3^{t-1}$, and the number of $\mathbf{d}$ such that $(\mathbf{d}\cdot\mathbf{g})=0$ is $3^{r-1}-1$. Since $\mathbf{g}\in U$, then the number of $\mathbf{d}$ satisfying  $wt((\mathbf{d}\cdot \mathbf{x})_{\mathbf{ x}\in U\setminus\{\mathbf{0}\}})=2\cdot3^{t-1}$ and $(\mathbf{d}\cdot\mathbf{g})=0$ is $3^{r-1}-3^{r-t}$. Thus $\mathcal{D}_{m,r}$ is a $[M+1,r,2a\cdot3^{r-1}]$ minimal code with the weight enumerator
\[W_{\mathcal D_{m,r}}(z)=1+(3^{r-t}-1)z^{2a3^{r-1}} +(3^{r-1}-3^{r-t})z^{2a3^{r-1}+2\cdot3^{t-1}}+2\cdot3^{r-1}z^{2a3^{r-1}+2\cdot3^{t-1}+2}.\]
Then, let $\mathcal{D}_{m,r}$ be the index code of $\mathcal{F}$, we have the following theorem by Theorem \ref{thm3}.
\begin{thm}\label{thm8}
Assume that $m\geq 3$ and $2\leq r\leq m$, with notations defined as above. Let $\mathcal{D}_{m,r}$ be the index code of $\mathcal{F}$, then $\mathcal{C}_\mathcal{F}$ of Theorem \ref{thm3} is a $[3^{2m}-1,2m+r,d]$ minimal code, and its weight enumerator is
\begin{eqnarray*}
W_{\mathcal{C}_{\mathcal{F}}}(z)=1+(M^2-1)z^{2M^2/3}+\sum_{w>0}A_w\left(z^{w(M-1)}+(M+1-w)(M-1)z^{\frac{2}{3}M^2-w} +w(M-1)z^{\frac{2}{3}M^2+M-w}\right),
\end{eqnarray*}
where
\[ d=\begin{cases}\frac{2}{3}M^2-2a3^{r-1}-2,&t=0,\\\frac{2}{3}M^2-2a3^{r-1}-2\cdot3^{t-1}-2,&1\leq t<r,\end{cases} \]
and
\[\begin{cases}(w,A_w)\in\{(2a3^{r-1},3^{r}-1),\ (2a3^{r-1}+2,2\cdot3^{r-1})\},&t=0,\\ (w,A_w)\in\{(2a3^{r-1},3^{r-t}-1),\ (2a3^{r-1}+2\cdot3^{t-1},3^{r-1}-3^{r-t}),\ (2a3^{r-1}+2\cdot3^{t-1}+2,2\cdot3^{r-1})\},&1\leq t<r.\end{cases}\]
Furthermore, the parameters for $\mathcal{C}_\mathcal{F}^\perp$ is $[3^{2m}-1,3^{2m}-1-2m-r,3]$, and the access structure of the secret-sharing scheme constructed from $\mathcal{C}_\mathcal{F}^\perp$ belongs to Type-II.
\end{thm}
\begin{proof}
The index code $\mathcal{D}_{m,r}$ is a minimal code with the parameters $[M+1,r, 2a\cdot3^{r-1}]$, and the maximum Hamming weight of $\mathcal{D}_{m,r}$ is
\[\begin{cases}
 2a\cdot3^{r-1}+2,&t=0,\\
 2a\cdot3^{r-1}+2\cdot3^{t-1}+2,&1\leq t<r.
 \end{cases}\]
Then, by $M+1=a(3^r-1)+(3^t-1)+2$, we have
\[2\leq 2a\cdot 3^{r-1}<2a\cdot3^{r-1}+2<2a\cdot3^{r-1}+2\cdot3^{t-1}+2< M-2.\]
The hypotheses of Theorem \ref{thm4} are therefore
satisfied, which proves $\mathcal{C}_\mathcal{F}$ is minimal. By Theorem \ref{thm3}, the explicit weight weight enumerator of $\mathcal{C}_\mathcal{F}$ follows from the weight enumerator of $\mathcal{D}_{m,r}$. From (\ref{shizi6}), we know the generator matrix of $\mathcal{D}_{m,r}$ has no zero columns. Thus, Proposition \ref{pro2} gives $d\bigl(\mathcal{C}_\mathcal{F}^\perp\bigr)=3$, and the access structure of the secret-sharing scheme based on $\mathcal{C}_\mathcal{F}^\perp$ is a Type-II access structure.

This completes the proof of Theorem \ref{thm8}.
\end{proof}
\begin{rem}
For convenience, put $u=2a\cdot3^{r-1}$, then
\[u-\frac{2M}{3}=2\frac{M-3^t}{3^r-1}3^{r-1}-\frac{2M}{3}=\frac{2(M-3^{r+t})}{3(3^r-1)}\geq0,\]
thus $u\geq \frac{2M}{3}$. From the weight weight enumerator of $\mathcal{C}_\mathcal{F}$, we know
\[w_{\max}=w_\ast(M-1),\qquad w_{\min}=\frac{2M^2}{3}-w_\ast.\]
where $w_\ast$ is the maximum Hamming weight of $\mathcal{D}_{m,r}$. Since $w_\ast\leq M-2$, therefore
\[\frac{w_{\min}}{w_{\max}}-\frac{2}{3}
 =\frac{2M^2-(2M+1)w_\ast}{3w_\ast(M-1)}
 \geq
 \frac{2M^2-(2M+1)(M-2)}{3w_\ast(M-1)}
 =\frac{3M+2}{3w_\ast(M-1)}>0.\]
Thus $\mathcal{C}_\mathcal{F}$ of Theorem \ref{thm8} satisfies the Ashikhmin-Barg condition.
\end{rem}
Based on Magma's program, the following example is presented, which is accordant with Theorem \ref{thm8}
\begin{exmp}
Let $m=5$ and $r=3$, then $t=2$ and $a=\frac{243-3^2}{3^3-1}=9$. The code $\mathcal{D}_{m,r}$ is a $[244,3,162]$ minimal code with the weight enumerator
 \[W_{\mathcal{D}_{m,r}}(z)=1+2z^{162}+6z^{168}+18z^{170}.\]
The code $\mathcal{C}_{\mathcal{F}}$ in Theorem \ref{thm8} is minimal with parameters $[59048,13,39196]$, and the weight enumerator is
\begin{eqnarray*}
W_{\mathcal{C}_{\mathcal{F}}}(z)&=&1+322344z^{39196}+110352z^{39198} +39690z^{39204}+59048z^{39366}\\
 &+&740520z^{39439}+243936z^{39441}+78408z^{39447}+6z^{40656}+18z^{41140}.
\end{eqnarray*}
Clearly, $\frac{w_{min}}{w_{max}}=\frac{39196}{41140}>\frac{2}{3}$. The parameters of the code $\mathcal{C}_{\mathcal{F}}^{\bot}$ in Theorem \ref{thm8} is $[59048,59035,3]$.
\end{exmp}

\subsection{$D$ is an infinite family derived from Krawtchouk polynomials}

In \cite{Heng-Ding-Zhou}, the authors constructed the ternary minimal code by
 \[\mathcal{C}_f=\{\mathbf{c}_{u,\mathbf{v}}=(uf(\mathbf{x})+\mathbf{v}\cdot \mathbf{x})_{\mathbf{x}\in {\mathbb{F}_3^l}^\ast}:u\in \mathbb{F}_3,\mathbf{v}\in\mathbb{F}_3^l\},\]
where $l$ is a positive integer. By choosing $f$ in a special way, the authors obtained the parameters of $\mathcal{C}_f$, as well as the following lemma.
\begin{lemma}\label{lem4}\cite{Heng-Ding-Zhou}
Let $l,k$ be integers with $l\geq 5$ and $2\leq k\leq \lfloor\frac{l-1}{2}\rfloor$. Then the linear code $\mathcal{C}_{f}$ is a minimal code with parameters
\[\left[3^l-1,l+1,\sum_{j=1}^k2^j\binom{l}{j}\right],\]
and the weight enumerator of $\mathcal{C}_{f}$ is given by
\[1+(3^l-1)z^{3^l-3^{l-1}}+2z^{\sum_{j=1}^k2^j\binom{l}{j}}+\sum_{i=1}^l2^{1+i}\binom{l}{i}z^{3^l-3^{l-1}+\Psi_k(i,l)-1},\]
where $\Psi_k(i,l)$ is the Lloyd polynomial.
\end{lemma}
In this subsection, we select the code $\mathcal{C}_f$ and extend it using (\ref{shizi3}) to obtain the index code $\mathcal{D}$. Based on (\ref{shizi3}), set $r=l+1$, one can obtain an $[M+1,r,\sum_{j=1}^k2^j\binom{r-1}{j}]$ minimal code $\mathcal{D}$, $\mathcal{D}$ and $\mathcal{C}_f$ have the same dimension and weight enumerator. Then, we have the following theorem by Theorem \ref{thm3}
\begin{thm}\label{thm6}
Suppose that $m$, $r$, $k$ are positive integers with $r\geq 6$, $r\leq m$, and $2\leq k\leq \lfloor\frac{r-2}{2}\rfloor$. Let $\mathcal{D}$ be the index code of $\mathcal{F}$ defined above, then $\mathcal{C}_\mathcal{F}$ of Theorem \ref{thm3} is a $[3^{2m}-1,2m+r,(3^m-1)\sum_{j=1}^k2^j\binom{r-1}{j}]$ minimal code violating the Ashikhmin-Barg condition, and its weight enumerator is
\begin{eqnarray*}
&&W_{\mathcal{C}_{\mathcal{F}}}(z)=1+(3^{2m}-1)z^{2\cdot3^{2m-1}}+2\left[z^{(3^m-1)\cdot\beta}+(3^{m}-1)\beta z^{2\cdot3^{2m-1}+3^m-\beta}+(3^m+1-\beta)(3^m-1)z^{2\cdot3^{2m-1}-\beta}\right] \\
&&+(3^{r-1}-1)\left[z^{2\cdot 3^{r-2}(3^m-1)}+2\cdot 3^{r-2}(3^{m}-1)z^{2\cdot3^{2m-1}+3^m-2\cdot 3^{r-2}}+(3^m+1-2\cdot 3^{r-2})(3^m-1)z^{2\cdot3^{2m-1}-2\cdot 3^{r-2}}\right]\\
&&+\sum_{i=1}^{r-1}2^{i+1}\binom{r-1}{i}\left[z^{\alpha\cdot (3^m-1)}+\alpha(3^{m}-1)z^{2\cdot3^{2m-1}+3^m-\alpha}+(3^m+1-\alpha)(3^m-1)z^{2\cdot3^{2m-1}-\alpha}\right],
\end{eqnarray*}
where $\beta=\sum_{j=1}^k2^j\binom{r-1}{j}$, $\alpha=2\cdot3^{r-2}+\Psi_k(i,r-1)-1$, and $\Psi_k(i,r-1)$ is the Lloyd polynomial.
Furthermore, the parameters for $\mathcal{C}_\mathcal{F}^\perp$ is $[3^{2m}-1,3^{2m}-1-2m-r,2]$, and the access structure of the secret-sharing scheme constructed from $\mathcal{C}_\mathcal{F}^\perp$ belongs to Type-I.
\end{thm}
\begin{proof}
The minimal code $\mathcal{D}$ has the same dimension and weight enumerator with that of $\mathcal{C}_f$ in Lemma \ref{lem4}, then $\mathcal{D}$ has minimum Hamming weight $\sum_{j=1}^k2^j\binom{r-1}{j}$ and maximum hamming weight $2\cdot3^{r-2}+2^k\binom{r-2}{k}-1$. Note that $m\geq r\geq 6$ and $2\leq k\leq \lfloor\frac{r-2}{2}\rfloor$, by Lemma \ref{lem3}, we have
\[2< \sum_{j=1}^k2^j\binom{r-1}{j}< 2\cdot3^{r-2}+\Psi_k(i,r-1)-1<2\cdot3^{r-2}+2^k\binom{r-2}{k}-1<2\cdot3^{r-2}+3^{r-2}-1< M-2.\]
Therefore, the conditions of Theorem \ref{thm4} are
satisfied, which proves $\mathcal{C}_\mathcal{F}$ is minimal. For nonzero codeword $\mathbf{d}\in \mathcal{D}$, by Theorem \ref{thm3}, the explicit weight weight enumerator of $\mathcal{C}_\mathcal{F}$ follows from the weight enumerator of $\mathcal{D}$. For convenience, put $\beta=\sum_{j=1}^k2^j\binom{r-1}{j}$, from the weight weight enumerator of $\mathcal{C}_\mathcal{F}$, we know
\[w_{\min}=\beta(M-1),\qquad w_{\max}=\frac{2M^2}{3}+M-\beta.\]
Then
\[\frac{w_{\min}}{w_{\max}}=\frac{\beta(M-1)}{\frac{2M^2}{3}+M-\beta},\]
the function $\frac{w_{\min}}{w_{\max}}$ is increasing for $\beta>0$, $\beta<3^{r-1}\leq \frac{M}{3}$ gives
\[\frac{w_{\min}}{w_{\max}}\leq\frac{(M/3)(M-1)}{2M^2/3+M-M/3}=\frac{M-1}{2(M+1)}<\frac12<\frac23.\]
Thus $\mathcal{C}_\mathcal{F}$ violates the Ashikhmin-Barg condition. Since the code length of $\mathcal{C}_f$ is $3^{r-1}-1<M+1$, (\ref{shizi3}) guarantees there exist at least one zero column in the generator matrix of $\mathcal{D}$, Proposition \ref{pro2} gives $d\bigl(\mathcal{C}_\mathcal{F}^\perp\bigr)=2$, and the access structure of the secret-sharing scheme based on $\mathcal{C}_\mathcal{F}^\perp$ is a Type-I access structure.

This completes the proof of Theorem \ref{thm6}.
\end{proof}

Based on Magma's program, the following example is presented, which is accordant with Theorem \ref{thm6}.
\begin{exmp}
Let $m=6$,$r=6$ and $k=2$, then $\mathcal{D}$ in Theorem \ref{thm6} is a $[730,6,50]$ minimal code with the weight enumerator
 \[W_{\mathcal{D}}(z)=1+2z^{50}+320z^{158}+242z^{162}+144z^{167}+20z^{185}.\]
The code $\mathcal{C}_{\mathcal{F}}$ in Theorem \ref{thm6} is minimal with parameters $[531440,18,36400]$, and the weight enumerator is
\begin{eqnarray*}
W_{\mathcal{C}_{\mathcal{F}}}(z)&=&1
 +2z^{36400}+320z^{115024}+242z^{117936}
 +144z^{121576}+20z^{134680}
+7\,935\,200z^{354109}
 +59\,020\,416z^{354127}\\
 &+&100\,067\,968z^{354132}
+133\,253\,120z^{354136}
 +990\,080z^{354244}
 +531\,440z^{354294}
+2\,693\,600z^{354838}\\
 &+&17\,506\,944z^{354856}
 +28\,540\,512z^{354861}
+36\,807\,680z^{354865}
 +72\,800z^{354973}.
\end{eqnarray*}
Clearly, $\frac{w_{min}}{w_{max}}={36400}{354973}\approx0.10254<\frac{2}{3}$. The parameters of the code $\mathcal{C}_{\mathcal{F}}^{\bot}$ in Theorem \ref{thm6} is $[531440,531422,2]$.
\end{exmp}
\subsection{$\mathcal{D}$ is an infinite family derived from cutting blocking sets}

Let $E=\{\mathbf{g}_0,\ldots,\mathbf{g}_{n-1} \}$ be subset of $\mathbb{F}_q^k$, and define the linear code $\mathcal{C}_E$ by
\[\mathcal{C}_E=\{\mathbf{c}(\mathbf{v})=(\mathbf{g}_0\cdot\mathbf{v},\ldots,\mathbf{g}_{n-1}\cdot \mathbf{v}):\mathbf{v}\in\mathbb{F}_q^k\}.\]
In \cite{Tang-Qi-Liao-Zhou}, let $h$ and $k$ be positive integers with $3\leq h\leq k$, based on the cutting blocking sets $E$, where $E$ is defined by
\[E=\{(x_1,\cdots,x_k)\in {\mathbb{F}_q^k}^\ast:x_1\cdots x_h=0\},\]
the authors proved the code $\mathcal{C}_E$ is minimal. And the parameters and weight enumerator of $\mathcal{C}_E$ are given by the following lemma, since we are interested in ternary minimal codes, the following lemma is stated for $q=3$.
\begin{lemma}\label{lem5}
Let $h$ and $k$ be two integers with $3\leq h\leq k$, then the code $\mathcal{C}_E$ is a $[(3^h-2^h)3^{k-h}-1,k,3^{k-h}(2\cdot3^{h-1}-2^h)]$ minimal linear code, and the weight enumerator of $\mathcal{C}_E$ is
\[1+(3^k-3^h)z^{2\cdot3^{k-h-1}(3^h-2^h)}+\sum_{s=1}^h2^s\binom{h}{s}z^{2\cdot3^{k-h-1}(3^h-2^h)+(-1)^s3^{k-h-1}2^{h-s+1}}.\]
\end{lemma}
In this subsection, we select the code $\mathcal{C}_E$ and extend it using (\ref{shizi3}) to obtain the index code $\mathcal{D}$. Based on (\ref{shizi3}), set $r=k$, one can obtain an $[M+1,r,3^{r-h}(2\cdot3^{h-1}-2^h)]$ minimal code $\mathcal{D}$, $\mathcal{D}$ and $\mathcal{C}_E$ have the same dimension and weight enumerator. Then, we have the following theorem by Theorem \ref{thm3}
\begin{thm}\label{thm7}
Suppose that $m$, $r$, $h$ are positive integers with $h\geq 3$, and $h\leq r< m$. Let $\mathcal{D}$ be the index code of $\mathcal{F}$ defined above, then $\mathcal{C}_\mathcal{F}$ of Theorem \ref{thm3} is a $[3^{2m}-1,2m+r,(3^m-1)3^{r-h}(2\cdot3^{h-1}-2^h)]$ minimal code violating the Ashikhmin-Barg condition, and its weight enumerator is
\begin{eqnarray*}
&&W_{\mathcal{C}_{\mathcal{F}}}(z)=1+(3^{2m}-1)z^{2\cdot3^{2m-1}}+(3^r-3^h)\left[z^{(3^m-1)\cdot\beta}+(3^{m}-1)\beta z^{2\cdot3^{2m-1}+3^m-\beta}+(3^m+1-\beta)(3^m-1)z^{2\cdot3^{2m-1}-\beta}\right] \\
&&+\sum_{s=1}^{h}2^{s}\binom{h}{s}\left[z^{\alpha\cdot (3^m-1)}+\alpha(3^{m}-1)z^{2\cdot3^{2m-1}+3^m-\alpha}+(3^m+1-\alpha)(3^m-1)z^{2\cdot3^{2m-1}-\alpha}\right],
\end{eqnarray*}
where $\beta=3^{r-h}(2\cdot3^{h-1}-2^h)$, $\alpha=2\cdot3^{r-h-1}(3^h-2^h)+(-1)^s3^{r-h-1}2^{h-s+1}$. Furthermore, the parameters for $\mathcal{C}_\mathcal{F}^\perp$ is $[3^{2m}-1,3^{2m}-1-2m-r,2]$, and the access structure of the secret-sharing scheme constructed from $\mathcal{C}_\mathcal{F}^\perp$ belongs to Type-I.
\end{thm}
\begin{proof}
The minimal code $\mathcal{D}$ has the same dimension and weight enumerator with that of $\mathcal{C}_E$ in Lemma \ref{lem5}, then $\mathcal{D}$ has minimum Hamming weight $3^{r-h}(2\cdot3^{h-1}-2^h)$ and maximum hamming weight $2\cdot3^{r-h-1}(3^h-2^h)+3^{r-h-1}2^{h-1}$.  By $m> r\geq h\geq 3$, we have
\[2<3^{r-h}(2\cdot3^{h-1}-2^h)< 2\cdot3^{r-h-1}(3^h-2^h)+3^{r-h-1}2^{h-1}=2\cdot3^{r-1}-3^{r-h}2^{h-1}<2\cdot3^{r-1}< M-2.\]
Therefore, the conditions of Theorem \ref{thm4} are
satisfied, which proves $\mathcal{C}_\mathcal{F}$ is minimal. For nonzero codeword $\mathbf{d}\in \mathcal{D}$, by Theorem \ref{thm3}, the explicit weight weight enumerator of $\mathcal{C}_\mathcal{F}$ follows from the weight enumerator of $\mathcal{D}$. For convenience, put $\beta=3^{r-h}(2\cdot3^{h-1}-2^h)$, from the weight weight enumerator of $\mathcal{C}_\mathcal{F}$, we know
\[ w_{\min}=\beta(M-1),\qquad w_{\max}=\frac{2M^2}{3}+M-\beta.\]
Then
\[\frac{w_{\min}}{w_{\max}}=\frac{\beta(M-1)}{\frac{2M^2}{3}+M-\beta},\]
the function $\frac{w_{\min}}{w_{\max}}$ is increasing for $\beta>0$, $\beta<2\cdot3^{r-1}< \frac{M}{3}$ gives
\[\frac{w_{\min}}{w_{\max}}\leq\frac{(M/3)(M-1)}{2M^2/3+M-M/3}=\frac{M-1}{2(M+1)}<\frac12<\frac23.\]
Thus $\mathcal{C}_\mathcal{F}$ violates the Ashikhmin-Barg condition. Since the code length of $\mathcal{C}_E$ is $3^{r-h}(3^h-2^h)-1<M+1$, (\ref{shizi3}) guarantees there exist at least one zero column in the generator matrix of $\mathcal{D}$, Proposition \ref{pro2} gives $d\bigl({\mathcal{C}_\mathcal{F}}^\perp\bigr)=2$, and the access structure of the secret-sharing scheme based on $\mathcal{C}_\mathcal{F}^\perp$ is a Type-I access structure.

This completes the proof of Theorem \ref{thm7}.
\end{proof}

Based on Magma's program, the following example is presented, which is accordant with Theorem \ref{thm7}.
\begin{exmp}
Let $m=4$ and $r=h=3$, then $\mathcal{D}$ in Theorem \ref{thm7} is a $[82,3,10]$ minimal code with the weight enumerator
 \[W_{\mathcal{D}}(z)=1+6z^{10}+8z^{12}+12z^{14}.\]
The code $\mathcal{C}_{\mathcal{F}}$ in Theorem \ref{thm7} is minimal with parameters $[6560,11,800]$, and the weight enumerator is
\begin{eqnarray*}
W_{\mathcal{C}_{\mathcal{F}}}(z)&=&1+6z^{800}+8z^{960}+12z^{1120}
+65\,280z^{4360}+44\,800z^{4362}+34\,560z^{4364}\\
&+&6\,560z^{4374}+13\,440z^{4441}
 +7\,680z^{4443}+4\,800z^{4445}.
\end{eqnarray*}
Clearly, $\frac{w_{min}}{w_{max}}=\frac{800}{4445}\approx0.17998<\frac{2}{3}$. The parameters of the code $\mathcal{C}_{\mathcal{F}}^{\bot}$ in Theorem \ref{thm7} is $[6560,6549,2]$.
\end{exmp}

The index code $\mathcal{D}$ in Theorems \ref{thm6} and \ref{thm7} is obtained by extension as in (\ref{shizi3}). Alternatively, it can also be extended as in (\ref{shizi6}) to obtain the index code $\mathcal{D}$ satisfying the conditions of Theorem \ref{thm4}, thereby yielding the minimal code $\mathcal{C}_{\mathcal{F}}$. However, with such an extension method, the weight enumerator of $\mathcal{D}$ is very difficult to determine, and consequently so is the weight enumerator of $\mathcal{C}_{\mathcal{F}}$. Therefore, we omit it here.

\begin{rem}
Apart from selecting the above three classes of minimal codes as index codes to construct the minimal code $\mathcal{C}_{\mathcal{F}}$ in Theorem \ref{thm3}, one may also select linear codes with a few weights satisfying the Ashikhmin-Barg condition as index codes to construct the minimal code $\mathcal{C}_{\mathcal{F}}$ of Theorem \ref{thm3}, and determine its weight enumerator. To avoid prolixity, we omit the details here.
\end{rem}

\section{A Class of Ternary Minimal Linear Codes from Krawtchouk Polynomials-Based Index Codes}

In this section, using the general construction (\ref{shizi2}), we shall present a class of ternary minimal codes with $w_{\min } / w_{\max } \leq 2 / 3$ by a class of specific functions, which are from index codes based on Krawtchouk polynomials. We always assume that $n\geq 1$ is a positive integer in this section, put $Q=3^n$, and define
\[L_j=\{\mathbf{x}\in\mathbb{F}_3^n:wt(\mathbf{x})=j\},\qquad 1\leq j\leq n,\]
then $L_1,\ldots,L_n$ form a partition of $\mathbb{F}_3^n\setminus\{0\}$.
Let $\mathcal{E}\subseteq \mathbb{F}_3^n$ be an $[n,r]$ ternary linear code, called the \emph{radial index code}, and let $\mathbf{e}_1, \ldots,\mathbf{e}_r$ be a basis of $\mathcal{E}$ over $\mathbb{F}_3$. For $\mathbf{e}_i=(e_{i,1},e_{i,2},\ldots,e_{i,n})$, $1\leq i\leq r$, define the function $f_i:\mathbb{F}_3^n\longrightarrow \mathbb{F}_3$ by
\begin{eqnarray}\label{shizi7}
f_i(\mathbf{x})=\left\{\begin{array}{ll}e_{i,j}, &\text{if}\quad \mathbf{x}\in L_j;\\
 0, &\text{otherwise}, \end{array}\right.
\end{eqnarray}
to indicate the relationship between $f_i$ and $\mathbf{e}_i$, we denote $f_i$ by $f_{\mathbf{e}_i}$. Based on $f_{\mathbf{e}_i}(1\leq i\leq r)$, the general construction (\ref{shizi2}) is used to obtain the ternary linear code $\mathcal{C}_{\mathcal{F}}$. And the weight enumerator of $\mathcal{C}_{\mathcal{F}}$ is given as follows according to the weight distribution of $\mathcal{E}$.
\begin{thm}\label{thm9}
Let notations defined as above, then the code $\mathcal{C}_{\mathcal{F}}$ in (\ref{shizi2}) is a $[3^n-1,n+r]$ ternary linear code, and the weight enumerator of $\mathcal{C}_{\mathcal{F}}$ is given as follows
\[W_{\mathcal{C}_{\mathcal{F}}}(z)=1+(Q-1)z^{\frac{2}{3}Q}+\sum_{\mathbf{e}\in \mathcal{E}\setminus \{\mathbf{0}\}}\left(z^{\sum_{j\in J_{\mathbf{e}}}2^j\binom{n}{j}}
 +\sum_{s=1}^{n}\binom ns2^s
 z^{\frac{2Q}{3}+\sum_{j\in J_{\mathbf{e}}}K_j(s,n)}\right),\]
where $J_{\mathbf{e}}=\text{Supp}(\mathbf{e})$.
\end{thm}
\begin{proof}
We first show the code $\mathcal{F}$ satisfies the assumption C1-C2, then the dimension of $\mathcal{C}_{\mathcal{F}}$ is $n+r$ by Theorem \ref{thm1}. According to (\ref{shizi7}), consider the map
\[\Psi:\mathcal{E}\longrightarrow \mathcal{F}, \mathbf{e}\longmapsto f_\mathbf{e}, \]
where $\mathcal{F}=\text{Span}_{\mathbb{F}_3}\{f_1,\ldots,f_r\}$, obviously, $\Psi$ is a bijection, and $\mathcal{F}$ satisfies the assumption C1. For $\mathbf{e}\neq \mathbf{0}$, choose $j\in \text{Supp}(\mathbf{e})$, let $\mathbf{x}\in L_j$, then $2\mathbf{x}\in L_j$, by (\ref{shizi7}), we have
\[f_{\mathbf{e}}(2\mathbf{v})=f_{\mathbf{e}}(\mathbf{v})\neq2f_{\mathbf{e}}(\mathbf{v}),\]
thus $f_{\mathbf{e}}$ is not linear, and $\mathcal{F}$ satisfies the assumption C2.

Next, we calculate the weight enumerator of $\mathcal{C}_{\mathcal{F}}$. By (\ref{shizi7}), for nonzero vector $\mathbf{e}=(e_1,\ldots,e_n)\in\mathcal{E}$, there exists $f_\mathbf{e}\in\mathcal{F}$ such that
\begin{eqnarray*}
f_\mathbf{e}(\mathbf{x})=\left\{\begin{array}{ll}e_j, &\text{if}\quad \mathbf{x}\in L_j;\\
 0, &\text{otherwise}. \end{array}\right.
\end{eqnarray*}
For $\mathbf{v}\in \mathbb{F}_3^n$, in order to obtain $wt(\mathbf{c}_{\mathbf{e}}(\mathbf{v}))$, where $\mathbf{c}_{\mathbf{e}}(\mathbf{v})=(f_\mathbf{e}(\mathbf{x})+\mathbf{x}\cdot\mathbf{v})_{\mathbf{x}\in{\mathbb{F}_3^n}^\ast}$, by Theorem \ref{thm1}, we need to compute $\widehat{f_\mathbf{e}(-\mathbf{v})}$. For convenience of computation, write $J_\mathbf{e}=\text{Supp}(\mathbf{e})$, and
\[S(\mathbf{e})=\sum_{j\in J_\mathbf{e}}(1-\zeta_3^{e_j}).\]
Then
$$
\begin{aligned}
\widehat{{f}_\mathbf{e}}(-\mathbf{v})&=\sum_{\mathbf{x} \in \mathbb{F}_3^n} \zeta_3^{f_{\mathbf{e}}(\mathbf{x})+\mathbf{v} \cdot \mathbf{x}}
=\sum_{j\in J_\mathbf{e}}\sum_{\mathbf{x}\in L_j}\zeta_3^{e_j+\mathbf{x}\cdot\mathbf{v}}
+\sum_{\mathbf{x}\in\mathbb{F}_3^n\setminus\bigcup_{j\in J_\mathbf{e}}L_j}\zeta_3^{\mathbf{x}\cdot\mathbf{v}}\\
&=\sum_{j\in J_\mathbf{e}}(\zeta_3^{e_j}-1)\sum_{\mathbf{x}\in
L_j}\zeta_3^{\mathbf{x}\cdot\mathbf{v}}+\sum_{\mathbf{x}\in\mathbb{F}_3^n}\zeta_3^{\mathbf{x}\cdot\mathbf{v}}.
\end{aligned}
$$
For $\mathbf{v} \in \mathbb{F}_3^m$, if $\mathbf{v}= \mathbf{0}$, then
\[\widehat{f_{\mathbf{e}}}(\mathbf{0})=Q-S(\mathbf{e})2^j\binom{n}{j},\]
for $e_j\in{\mathbb{F}_3}^\ast$, $Re(\zeta_3^{e_j}-1)=-\frac{3}{2}$, thus
\[Re(\widehat{f_{\mathbf{e}}}(\mathbf{0}))=Q-\frac{3}{2}\sum_{j\in J_\mathbf{e}}2^j\binom{n}{j}.\]
If $\mathbf{v}\neq \mathbf{0}$, let $wt(\mathbf{v})=s(1\leq s\leq n)$, the number of $\mathbf{v}$ is $2^s\binom{n}{s}$, then
\[\widehat{f_{\mathbf{e}}}(-\mathbf{v})=-S(\mathbf{e})K_j(s,n),\]
and
\[Re(\widehat{f_{\mathbf{e}}}(-\mathbf{v}))=-\frac{3}{2}\sum_{j\in J_\mathbf{e}}K_j(s,n).\]
The weight of the code $\mathbf{c}_{\mathbf{e}}(\mathbf{v})$ follow from Theorem \ref{thm1}.

If $\mathbf{e}=\mathbf{0}\in\mathcal{E}$, then $\mathbf{c}_{\mathbf{0}}(\mathbf{v})=(\mathbf{v}\cdot\mathbf{x})_{\mathbf{x}\in{\mathbb{F}_3^n}^\ast}$, and $wt(\mathbf{c}_{\mathbf{0}}(\mathbf{v}))=\frac{2}{3}Q$ for $\mathbf{v}\neq \mathbf{0}$. The number of $\mathbf{v}\neq \mathbf{0}$ is $Q-1$.

This completes the proof of Theorem \ref{thm9}.
\end{proof}
Comparing Theorem \ref{thm3} with Theorem \ref{thm9}, we see that to determine the weight enumerator of $\mathcal{C}_{\mathcal{F}}$, Theorem \ref{thm3} requires only the weight distribution of the index code $\mathcal{D}$, whereas Theorem \ref{thm9} requires the exact support of each codeword in the index code $\mathcal{E}$. Determining the exact support is a difficult even for simplex codes, consequently, determining the weight enumerator in Theorem \ref{thm9} is difficult as well.

To prove that $\mathcal{C}_{\mathcal{F}}$ is minimal, we need some properties of $L_j$, which are stated in the following lemma.
\begin{lemma}\label{lem6}
Let $n$ and $j$ be positive integers with $n\geq3$ and $1\leq j\leq n$, and put
\[L_j=\{\mathbf{x}\in\mathbb{F}_3^n:wt(\mathbf{x})=j\},\]
then

(1) $\text{Span}_{\mathbb{F}_3}\{\mathbf{x}|\mathbf{x}\in L_j\}=\mathbb{F}_3^n$,

(2) for every $\mathbf{0}\neq \mathbf{w}\in\mathbb{F}_3^n$ and $2\leq j\leq n-1$,
 \[\{\mathbf{w}\cdot \mathbf{x}:\mathbf{x}\in L_j\}=\mathbb{F}_3, \]

(3) for every $\mathbf{0}\neq \mathbf{w}\in\mathbb{F}_3^n$, at least one of the two endpoint layers satisfies
\[\{\mathbf{w}\cdot\mathbf{ x}:\mathbf{x}\in L_1\}=\mathbb{F}_3\qquad\text{or}\qquad\{\mathbf{w}\cdot \mathbf{x}:\mathbf{x}\in L_n\}=\mathbb{F}_3.\]
\end{lemma}

\begin{proof}
We first prove the first assertion, to prove that $L_j$ spans $\mathbb{F}_3^n$, we will show that the standard
basis of $\mathbb{F}_3^n$ are contained in $\text{Span}_{\mathbb{F}_3}\{\mathbf{x}|\mathbf{x}\in L_j\}$. Let $\mathbf{a}_i(i=1,2,\ldots,n)$ be the standard basis of $\mathbb{F}_3^n$, for any $i\in\{1,\ldots,n\}$, since $j\leq n$, there exists a set
$T\subseteq\{1,\ldots,n\}\setminus\{i\}$ with size $j-1$, define
\[\mathbf{x}=\mathbf{a}_i+\sum_{t\in T}\mathbf{a}_t,\qquad \mathbf{y}=2\mathbf{a}_i+\sum_{t\in T}\mathbf{a}_t.\]
Then $\mathbf{x},\mathbf{y}\in L_j$, and $\mathbf{a}_i=\mathbf{y}-\mathbf{x}\in\text{Span}_{\mathbb{F}_3}\{\mathbf{x}:\mathbf{x}\in L_j\}$. Thus $L_j$ spans $\mathbb{F}_3^n$.

For the second assertion, fix $\mathbf{0}\neq \mathbf{w}=(w_1,\cdots,w_n)\in\mathbb{F}_3^n$ and write $t=wt(\mathbf{w})$. We first construct
$\mathbf{x}\in L_j$ such that $\mathbf{w}\cdot \mathbf{x}=0$ depending on the value of $t$. If $t=1$, the complement of $\text{Supp}(\mathbf{w})$ has size $n-1$. Since $j\leq n-1$, choose all $j$ positions of $\text{Supp}(\mathbf{x})$ in that complement and assign arbitrary nonzero entries to them, then $wt(\mathbf{x})=j$ and $\mathbf{w}\cdot \mathbf{x}=0$. Now suppose $t\geq2$, put $k=\min(j,t)$, choose $k$ positions in $\text{Supp}(\mathbf{w})$.
If $k$ is even, partition $k$ positions into $\{i_1,\cdots,i_{\frac{k}{2}}\}$ and $\{i_{1+\frac{k}{2}},\cdots,i_k\}$, choose the components $(x_{i_l},x_{i_{l+\frac{k}{2}}})$ of $\mathbf{x}$, $l=1,\cdots,\frac{k}{2}$, such that $w_{i_l}x_{i_l}=1$ and $w_{i_{l+\frac{k}{2}}}x_{i_{l+\frac{k}{2}}}=2$. If $k$ is odd, then $k\geq3$, partition $k$ positions into $\{i_1,\cdots,i_{\frac{k-3}{2}}\}$, $\{i_{\frac{k-3}{2}+1},\cdots,i_{k-3}\}$ and $\{i_{k-2},i_{k-1},i_k\}$, choose the components $(x_{i_l},x_{i_{l+\frac{k-3}{2}}})$ and $(x_{i_{k-2}},x_{i_{k-1}},x_{i_k})$ of $\mathbf{x}$, $l=1,\cdots,\frac{k-3}{2}$, such that $w_{i_l}x_{i_l}=1$, $w_{i_{l+\frac{k-3}{2}}}x_{i_{l+\frac{k-3}{2}}}=2$ and $w_{i_{k-2}}x_{i_{k-2}}=w_{i_{k-1}}x_{i_{k-1}}=w_{i_{k}}x_{i_{k}}=1$. Thus the entries chosen on the $k$ positions in $\text{Supp}(\mathbf{w})$ have total dot-product contribution zero. If $j\leq t$, then $k=j$, so these positions already form the support of a vector $x\in L_j$. If $j>t$, then $k=t$, assign arbitrary nonzero entries on
the $j-t$ positions from the complement of $\text{Supp}(\mathbf{w})$. Therefore, $\mathbf{w}\cdot \mathbf{x}=0$, and $\mathbf{x}\in L_j$.
By the first assertion that $L_j$ spans $\mathbb{F}_3^n$, if $\mathbf{w}\cdot \mathbf{x}=0$ for every $\mathbf{x}\in L_j$, then the linear function $\mathbf{w}\cdot \mathbf{x}$ would vanish on all of $\mathbb{F}_3^n$, which leads to $\mathbf{w}=0$ a contradiction. Hence there exists some $\mathbf{y}\in L_j$ such that $\mathbf{w}\cdot \mathbf{y}=a\neq0$. Since $2\mathbf{y}\in L_j$, and $\mathbf{w}\cdot(2\mathbf{y})=2a$, the elements $a$ and $2a$ are the two nonzero elements of $\mathbb{F}_3$. Based on discussions above, we have $\{\mathbf{w}\cdot \mathbf{x}:\mathbf{x}\in L_j\}=\mathbb{F}_3$ for $2\leq j\leq n-1$.

We prove the final assertion by considering two cases. We first suppose that $w_i=0$ for some $i$, we can choose $\mathbf{x}\in L_1$ with $x_i=1$, then $\mathbf{w}\cdot \mathbf{x}=0$. Since $L_1$ spans $\mathbb{F}_3^n$, we get $\{\mathbf{w}\cdot\mathbf{ x}|\mathbf{x}\in L_1\}=\mathbb{F}_3$ by a similar argument to the proof of assertion (2). Now suppose that $\mathbf{w}$ has full support, analogously to the proof of Assertion (2), the components of $\mathbf{x}$ are chosen according to the parity of $n$, this yields there exists $\mathbf{x}\in L_n$ such that $\mathbf{x}\cdot \mathbf{w}=0$. Since $L_n$ spans $\mathbb{F}_3^n$, we get $\{\mathbf{w}\cdot\mathbf{ x}|\mathbf{x}\in L_n\}=\mathbb{F}_3$ by a similar argument to the proof of assertion (2). Thus assertion (3) holds.

This completes the proof of Lemma \ref{lem6}.
\end{proof}

Next, based on Theorem \ref{thm2}, we will show the condition for $\mathcal{C}_{\mathcal{F}}$ to be a minimal code in Theorem \ref{thm9} from the perspective of the index code $\mathcal{E}$.
\begin{thm}\label{thm10}
Suppose that $S\subseteq\{1,2,\ldots,n\}$ with $|S|=h\geq4$, and let
$\mathcal{E}$ be an $[n,r]$ index code supported on $S$. Define
\[\Omega_n^{\max}(\mathcal{E})=\max_{\mathbf{e}\in\mathcal{E}}\sum_{j\in J_{\mathbf{e}}}2^j\binom nj.\]
If $\mathcal{E}$ satisfies the following three conditions simultaneously,

(B1) the index code $\mathcal{E}$ is a minimal ternary code,

(B2) every nonzero $\mathbf{e}\in\mathcal{E}$ satisfies $2\leq wt(\mathbf{e})\leq h-2$,

(B3) $\Omega_n^{\max}(\mathcal{E})<2Q/15$,

then the code $\mathcal{C}_{\mathcal{F}}$ in Theorem \ref{thm9} is minimal.
\end{thm}
\begin{proof}
Let $\mathbf{c}_1$ and $\mathbf{c}_2\in\mathcal{C}_{\mathcal{F}}$ be arbitrary nonzero codewords, for convenience, denote
\[\mathbf{c}_i=(f_{\mathbf{e}_i}(\mathbf{x})+\mathbf{v}_i\cdot \mathbf{x})_{\mathbf{x} \in {\mathbb{F}_3^n}^\ast}\
\text{and}\ J_{\mathbf{e}_i}=\text{Supp}(\mathbf{e}_i),\]
where $\mathbf{e}_i=(e_{i,1},\ldots,e_{i,n})\in \mathcal{E}$, $i=1,2$.
Assuming that $\text{Supp}(\mathbf{c}_1)\subseteq \text{Supp}(\mathbf{c}_2)$, by the conditions B1-B3, we prove that either $\mathbf{c}_1$ and $\mathbf{c}_2$ are proportional or the assumption does not hold. According to the relation between $\mathbf{v}_1$ and $\mathbf{v}_2$,
we shall discuss this by considering the following four cases.

{\bf Case 1}. $\mathbf{v}_1=\mathbf{v}_2=\mathbf{0}$. Then $\text{Supp}(\mathbf{c}_1)\subseteq \text{Supp}(\mathbf{c}_2)$ is equivalent to
$J_{\mathbf{e}_1}\subseteq J_{\mathbf{e}_2}$. Since $\mathbf{c}_1$ and $\mathbf{c}_2$ are nonzero, then the index words $\mathbf{e}_1$ and $\mathbf{e}_2$ are nonzero, and the minimality of $\mathcal{E}$ of the condition B1 implies
$\mathbf{e}_1=\lambda\mathbf{e}_2$ for some $\lambda\in\mathbb{F}_3^\ast$, so $\mathbf{c}_1=\lambda \mathbf{c}_2$.

{\bf Case 2}. Exactly one of $\mathbf{v}_i(i=1,2)$ is zero. We first assume that $\mathbf{v}_1=\mathbf{0}$ and $\mathbf{v}_2\neq\mathbf{0}$,  then $\mathbf{e}_1\neq \mathbf{0}$, the condition B2 gives $wt(\mathbf{e}_1)\geq2$, hence there is some $j\in \text{Supp}(\mathbf{e}_1)\setminus\{1,n\}$, otherwise $j\in\{1,n\}$. For either of the two cases, the assertions (2) and (3) of Lemma \ref{lem6} give that there exists $\mathbf{x}\in L_j$ satisfying $e_{2,j}+\mathbf{v}_2\cdot \mathbf{x}=0$, but $e_{1,j}\neq0$, which contradicts to $\text{Supp}(\mathbf{c}_1)\subseteq \text{Supp}(\mathbf{c}_2)$. Conversely, suppose that $\mathbf{v}_1\neq\mathbf{0}$ and $\mathbf{v}_2=\mathbf{0}$, then $\mathbf{e}_2\neq \mathbf{0}$, the bound $wt(\mathbf{e}_2)\leq h-2$ in the condition B2 leaves at least two elements of
$S\setminus \text{Supp}(\mathbf{e}_2)$, so one can choose $j\in S\setminus \text{Supp}(\mathbf{e}_2)$, then $\mathbf{c}_2$ vanishes on all $\mathbf{x}\in L_j$. So $\text{Supp}(\mathbf{c}_1)\subseteq \text{Supp}(\mathbf{c}_2)$ forces
\[e_{1,j}+\mathbf{v}_1\cdot \mathbf{x}=0\quad \text{and}\quad e_{1,j}+\mathbf{v}_1\cdot 2\mathbf{x}=0 \]
for all $\mathbf{x}\in L_j$. Comparing this equation at $\mathbf{x}$ and $2\mathbf{x}$ gives $e_{1,j}=0$ and $\mathbf{v}_1\cdot \mathbf{x}=0$ throughout $L_j$, then $\mathbf{v}_1=\mathbf{0}$ contradicts to the assumption $\mathbf{v}_1\neq\mathbf{0}$.

{\bf Case 3}. $\mathbf{v}_1,\mathbf{v}_2$ are nonzero and linearly dependent. There exists a $\lambda\in{\mathbb{F}_3}^\ast$ such that $\mathbf{v}_2=\lambda \mathbf{v}_1$, and $\lambda \mathbf{c}_1=(f_{\lambda\mathbf{e}_1}(\mathbf{x})+\lambda\mathbf{v}_1\cdot \mathbf{x})_{\mathbf{x} \in {\mathbb{F}_3^n}^\ast}=(f_{\lambda\mathbf{e}_1}(\mathbf{x})+\mathbf{v}_2\cdot \mathbf{x})_{\mathbf{x} \in {\mathbb{F}_3^n}^\ast}$. On every selected $L_j$ for $2\leq j\leq n-1$, by (2) of Lemma \ref{lem6}, the equation $e_{2,j}+\mathbf{v}_2\cdot \mathbf{x}=0$ always has a solution $\mathbf{x}$. At such an $\mathbf{x}$, $\text{Supp}(\mathbf{c}_1)\subseteq \text{Supp}(\mathbf{c}_2)$ forces $\lambda e_{1,j}+\mathbf{v}_2\cdot \mathbf{x}=0$, hence $\lambda e_{1,j}=e_{2,j}$ for $2\leq j\leq n-1$. Consequently $\lambda\mathbf{e}_1-\mathbf{e}_2\in\mathcal{E}$ and $\text{Supp}(\lambda\mathbf{e}_1-\mathbf{e}_2)\subseteq\{1,n\}$. If $\lambda\mathbf{e}_1\neq\mathbf{e}_2$, the condition B2 gives $\text{Supp}(\lambda\mathbf{e}_1-\mathbf{e}_2)=\{1,n\}$, by (3) of Lemma \ref{lem6}, choose
$j\in\{1,n\}$ such that $\{\mathbf{v}_2\cdot \mathbf{x}:\mathbf{x}\in L_j\}=\mathbb{F}_3$, then there exists an $\mathbf{x}\in L_j$ with
$e_{2,j}+\mathbf{v}_2\cdot \mathbf{x}=0$. $\text{Supp}(\mathbf{c}_1)\subseteq \text{Supp}(\mathbf{c}_2)$ gives
$\lambda e_{1,j}+\mathbf{v}_2\cdot \mathbf{x}=0$, hence
$\lambda e_{1,j}=e_{2,j}$. This contradicts to
$j\in \text{Supp}(\lambda\mathbf{e}_1-\mathbf{e}_2)$. Therefore
$\mathbf{e}_2=\lambda\mathbf{e}_1$, $\mathbf{c}_1$ and $\mathbf{c}_2$ are proportional.

{\bf Case 4}. $\mathbf{v}_1,\mathbf{v}_2$ are linearly independent. Then $\mathbf{v}_1,\mathbf{v}_2,\mathbf{v}_2+\mathbf{v}_1$, and $\mathbf{v}_2-\mathbf{v}_1$ are all nonzero. From the proof of Theorem \ref{thm9}, we know $Re(f_{\mathbf{e}}(\mathbf{v}))=-\frac{3}{2}\sum_{j\in J_\mathbf{e}}K_j(s,n)$, where $s=wt(\mathbf{v})$, since $|K_j(s,n)|\leq |L_j|=2^j\binom nj$, then
\[|Re(f_{\mathbf{e}}(\mathbf{v}))|=\frac{3}{2}\left|\sum_{j\in J_{\mathbf{e}}}K_j(s,n)\right|
 \leq\frac{3}{2}\sum_{j\in J_{\mathbf{e}}}2^j\binom nj
 \leq\frac{3}{2}\Omega_n^{\max}(\mathcal{E}).\]
If $\text{Supp}(\mathbf{c}_1)\subseteq \text{Supp}(\mathbf{c}_2)$, Theorem \ref{thm2} would make the Walsh containment defect zero. However, by condition B3, we have
\begin{eqnarray*}
 &&|Re_{f_{\mathbf{e}_2+\mathbf{e}_1}}(-\mathbf{v}_2-\mathbf{v}_1)
     +Re_{f_{\mathbf{e}_2-\mathbf{e}_1}}(-\mathbf{v}_2+\mathbf{v}_1)
 -2Re_{f_{\mathbf{e}_2}}(-\mathbf{v}_2)+Re_{f_{\mathbf{e}_1}}(-\mathbf{v}_1)|\\
 &&\leq|Re_{f_{\mathbf{e}_2+\mathbf{e}_1}}(-\mathbf{v}_2-\mathbf{v}_1)|
     +|Re_{f_{\mathbf{e}_2-\mathbf{e}_1}}(-\mathbf{v}_2+\mathbf{v}_1)|
 +2|Re_{f_{\mathbf{e}_2}}(-\mathbf{v}_2)|+|Re_{f_{-\mathbf{e}_1}}(\mathbf{v}_1)|\\
 &&\leq(1+1+2+1)\frac{3}{2}\Omega_n^{\max}(\mathcal{E})=\frac{15}{2}\Omega_n^{\max}(\mathcal{E})<Q.
\end{eqnarray*}
This is a contradiction to Theorem \ref{thm2}.

From the above discussion, for any two nonzero codewords $\mathbf{c}_1$ and $\mathbf{c}_2$, the assumption $\text{Supp}(\mathbf{c}_1)\subseteq\text{Supp}(\mathbf{c}_2)$ forces that
either $\mathbf{c}_1$ and $\mathbf{c}_2$ are proportional or the assumption does not hold, thus $\mathcal{C}_\mathcal{F}$ is minimal.

This completes the proof of Theorem \ref{thm10}.
\end{proof}
Theorem \ref{thm10} requires $S$ to be the support of $\mathcal{E}$. In fact, this requirement can be relaxed to $\text{Supp}(\mathcal{E})\subseteq S$, where $\text{Supp}(\mathcal{E})=\bigcup_{\mathbf{e}\in\mathcal{E}}\text{Supp}(\mathbf{e})$, and Theorem \ref{thm10} remains valid.
\begin{rem}\label{rem3}
Denote by $w_{\max}$ and $w_{\min}$ the maximum and minimum Hamming weights of $\mathcal{C}_\mathcal{F}$, respectively. If the index code $\mathcal{E}$ of $\mathcal{C}_\mathcal{F}$ satisfies the conditions of Theorem \ref{thm10}, then by Theorem \ref{thm9}, it follows that
\[0<w_{\min}\leq\Omega_n^{\max}(\mathcal{E})<\frac{2\cdot3^n}{15},
 \qquad
 w_{\max}\geq\frac{2\cdot3^n}{3},
 \qquad
 \frac{w_{\min}}{w_{\max}}<\frac{\frac{2\cdot3^n}{15}}{\frac{2\cdot3^n}{3}}=\frac{1}{5}<\frac{2}{3}.\]
Therefore, $\mathcal{C}_\mathcal{F}$ of \ref{thm10} violates the Ashikhmin-Barg condition.
\end{rem}

To determine the access structure of secret sharing schemes based on $\mathcal{C}_{\mathcal{F}}^\bot$, we need to consider the minimum distance of $\mathcal{C}_{\mathcal{F}}^\bot$, the minimum distance of $\mathcal{C}_{\mathcal{F}}^\bot$ is determined by the generator matrix of $\mathcal{E}$.
\begin{pro}\label{pro3}
Assume that $n\geq4$, let $B$ be an $r\times n$ generator matrix of the index code $\mathcal{E}$, and let $\mathbf{b}_j$ denote
$j$th column of $B$ with $1\leq j\leq n$, then the distance of $\mathcal{C}_\mathcal{F}^\perp$ is
\begin{equation*}
 d\bigl(\mathcal{C}_\mathcal{F}^\perp\bigr)
 =\begin{cases}
 2,&\mathbf{b}_j=0\text{ for some $j$,}\\
 3,&\mathbf{b}_j\neq0\text{ for every $j$}.
 \end{cases}
\end{equation*}
\end{pro}
\begin{proof}
Treat the chosen rows of $B$ as a basis of $\mathcal{E}$, for $1\leq j\leq n$, by (\ref{shizi7}), the column of the generator matrix of
$\mathcal{C}_{\mathcal{F}}$ indexed by $\mathbf{x}\in L_j$ is
\[g_\mathbf{x}=\binom{\mathbf{b}_j}{\mathbf{x}}\in\mathbb{F}_3^{r+n}.\]
No such column is zero because of $\mathbf{x}\neq\mathbf{0}$, so $d\bigl({\mathcal{C}_\mathcal{F}}^\perp\bigr)\geq 2$.

Suppose two distinct columns $g_\mathbf{x}$ and $g_\mathbf{y}$ are proportional. Their last $n$ entries give
$\mathbf{y}=\lambda \mathbf{x}$ for some $\lambda\in{\mathbb{F}_3}^\ast$, so $\mathbf{y}\in L_j$ for the same $j$. Since $\mathbf{y}$ and $\mathbf{x}$ are distinct, then $\lambda=2$ and
$\mathbf{y}=2\mathbf{x}$. The first $r$ entries then require $\mathbf{b}_j=2\mathbf{b}_j$, equivalently $\mathbf{b}_j=\mathbf{0}$. Conversely,
if $\mathbf{b}_j=\mathbf{0}$, then $F(\mathbf{x})=F(2\mathbf{x})=0$
for every $\mathbf{x}\in L_j$, where $\mathbf{F}(\mathbf{x})=(f_1(\mathbf{x}),\ldots,f_r(\mathbf{x}))^T$. Thus, by Proposition
\ref{Pro1}, the first case of Proposition \ref{pro3} holds.

Now assume that every $\mathbf{b}_j$ is nonzero. The preceding argument shows that no two columns are
proportional, so $d\bigl(\mathcal{C}_\mathcal{F}^\perp\bigr)\geq 3$. Since $n\geq4$, let
\[\mathbf{x}=(1,1,1,0,0,\ldots,0),\
 \mathbf{y}=(1,2,0,1,0,\ldots,0),\
 \mathbf{z}=(1,0,2,2,0,\ldots,0),\]
and $\mathbf{x},\mathbf{y},\mathbf{z}\in L_3$, then we have
\[g_\mathbf{x}+g_\mathbf{y}+g_\mathbf{z}=\binom{3\mathbf{b}_3}{\mathbf{x}+\mathbf{y}+\mathbf{z}}=\mathbf{0}.\]
Hence $d\bigl({\mathcal{C}_\mathcal{F}}^\perp\bigr)=3$.

This completes the proof of Proposition \ref{pro2}.
\end{proof}

\begin{rem}
The construction in Theorem \ref{thm9} is general. Provided that the index code $\mathcal{E}$ satisfying the conditions of Theorem \ref{thm10}, the minimal code ${\mathcal{C}_\mathcal{F}}$ can be obtained via Theorem \ref{thm9}. However, owing to the properties of $L_j$, it is difficult to determine the weight enumerator of ${\mathcal{C}_\mathcal{F}}$ exactly, since this requires precise knowledge of the support of each codeword in $\mathcal{E}$.
\end{rem}

In what follows, we choose a projective simplex code, extend it to obtain the index code $\mathcal{E}$ satisfying the conditions of Theorem \ref{thm10}, and then construct the ternary minimal ${\mathcal{C}_\mathcal{F}}$.

Assume that $r$ and $h$ are positive integers with $r\geq2$ and $h\geq \frac{3^r-1}{2}$, let $\mathcal{E}_r$ be an $[\frac{3^r-1}{2},r,3^{r-1}]$ projective simplex code. By adding zero components, we extend $\mathcal{E}_r$ to the code $\mathcal{E}_h$ of length $h$. Choose an ordered set $S=(j_1,\ldots,j_h)\subseteq\{1,\ldots,n\}$, and embed $\mathcal{E}_h$ into $\mathbb{F}_3^n$ indexed by $S$ by setting
\[\mathcal{E}=\left\{\mathbf{e}=(e_1,e_2,\ldots,e_n)\in\mathbb{F}_3^n:(e_{j_1},\ldots,e_{j_h})\in \mathcal{E}_h,\
 e_j=0\text{ for $j\notin S$}\right\}.\]
Then $\mathcal{E}$ and $\mathcal{E}_r$ have the same dimension and weight enumerator. Using $\mathcal{E}$ as the index code of ${\mathcal{C}_\mathcal{F}}$, we apply Theorem \ref{thm9} to obtain the following theorem.
\begin{thm}\label{thm11}
Suppose that $r$, $h$, $n$ are positive integers with $r\geq 2$, and $\frac{3^r-1}{2}<h\leq n$. Let $\mathcal{E}$ be the index code of $\mathcal{F}$ defined above, if $\Omega_n^{\max}(\mathcal{E})<\frac{2Q}{15}$, then $\mathcal{C}_\mathcal{F}$ of Theorem \ref{thm9} is a $[3^{n}-1,n+r,\min_{\mathbf{e}\in\mathcal{E}\backslash\{\mathbf{0}\}}\sum_{j\in J_{\mathbf{e}}}2^j\binom nj]$ ternary minimal code violating the Ashikhmin-Barg condition, and its weight enumerator is
\[W_{\mathcal{C}_{\mathcal{F}}}(z)=1+(Q-1)z^{\frac{2}{3}Q}+\sum_{\mathbf{e}\in \mathcal{E}\setminus \{\mathbf{0}\}}\left(z^{\sum_{j\in J_{\mathbf{e}}}2^j\binom{n}{j}}
 +\sum_{s=1}^{n}\binom ns2^s
 z^{\frac{2Q}{3}+\sum_{j\in J_{\mathbf{e}}}K_j(s,n)}\right),\]
where $J_{\mathbf{e}}=\text{Supp}(\mathbf{e})$. Furthermore, the parameters for $\mathcal{C}_\mathcal{F}^\perp$ is $[3^{n}-1,3^{n}-1-n-r,2]$, and the access structure of the secret-sharing scheme constructed from $\mathcal{C}_\mathcal{F}^\perp$ belongs to Type-I.
\end{thm}
\begin{proof}
Since $\mathcal{E}_r$ is minimal, by the definition of $\mathcal{E}$, we know $\mathcal{E}$ is minimal satisfying the condition B1 of Theorem \ref{thm10}. It follows from the definition of $\mathcal{E}$ that $|\text{Supp}(\mathcal{E})|=\frac{3^r-1}{2}$ and $wt(\mathbf{e})=3^{r-1}$ for $\mathbf{e}\in \mathcal{E}$, consequently,
\[h-2-wt(\mathbf{e})\geq 1+\frac{3^r-1}{2}-2-3^{r-1}=\frac{3^{r-1}-3}{2}\geq0\ \text{and} \ wt(\mathbf{e})=3^{r-1}\geq 2,\]
due to $r\geq 2$, so $\mathcal{E}$ satisfies the condition B2 in Theorem \ref{thm10}. Moreover, by the assumption $\Omega_n^{\max}(\mathcal{E})<\frac{2Q}{15}$, thus $\mathcal{E}$ satisfies the conditions of Theorem \ref{thm10}, hence $\mathcal{C}_\mathcal{F}$ is minimal. The weight weight enumerator of $\mathcal{C}_\mathcal{F}$ follows from Theorem \ref{thm9}, and the minimal Hamming distance of $\mathcal{C}_\mathcal{F}$ is $\min_{\mathbf{e}\in\mathcal{E}}\sum_{j\in J_{\mathbf{e}}}2^j\binom nj$. Remark \ref{rem3} shows that $\mathcal{C}_\mathcal{F}$ violates the Ashikhmin-Barg condition.

Since $h>\frac{3^r-1}{2}$, adding zero components guarantees there exist at least one zero column in the generator matrix of $\mathcal{E}$, Proposition \ref{pro3} gives $d\bigl(\mathcal{C}_\mathcal{F}^\perp\bigr)=2$, and the access structure of the secret-sharing scheme based on $\mathcal{C}_\mathcal{F}^\perp$ is a Type-I access structure.

This completes the proof of Theorem \ref{thm11}.
\end{proof}
\begin{rem}
Here $\mathcal{E}$ is taken to be an extended projective simplex code. As in the previous chapter, one may also choose minimal codes of other types and extend them by means of (\ref{shizi3}) and (\ref{shizi6}) to obtain the index code $\mathcal{E}$ satisfying the conditions of Theorem \ref{thm10}, thereby constructing the $\mathcal{C}_\mathcal{F}$ of Theorem \ref{thm9}. The details are omitted here for the sake of brevity.
\end{rem}
Based on Magma's program, the following example is presented, which is accordant with Theorem \ref{thm11}.
\begin{exmp}
Let $r=2$, $h=5$ and $n=8$, then $\mathcal{E}_h$ is a $[5,2,3]$ minimal code with the generator matrix
 \[G_h=
 \begin{pmatrix}
  1&1&0&0&1\\
  0&1&1&0&2
 \end{pmatrix}.\]
We extend $\mathcal{E}_h$ to the code $\mathcal{E}$ with the generator matrix
\[ G_{\mathcal{E}}
 =\begin{pmatrix}
  1&1&0&0&0&0&0&1\\
  0&1&1&0&0&0&0&2
 \end{pmatrix},\]
thus $\text{Supp}(\mathcal{E})=\{1,2,3,8\}$. For $\mathbf{e}\in \mathcal{E}$, $J_\mathbf{e}\in\bigl\{\{1,2,8\},\ \{2,3,8\},\ \{1,2,3\},\ \{1,3,8\}\bigr\}$, $2\leq wt(\mathbf{e})=3\leq h-2=3$, $\Omega_n^{\max}(\mathcal{E})=4\binom{8}{2}+8\binom{8}{3}+2^8\binom{8}{8}=816<\frac{2\cdot3^8}{15}$, and $d=\min_{\mathbf{e}\in\mathcal{E}}\sum_{j\in J_{\mathbf{e}}}2^j\binom nj=2\binom81+4\binom82+2^8\binom88=384$.
The code $\mathcal{C}_{\mathcal{F}}$ of Theorem \ref{thm11} is a minimal code with parameters $[6560, 10, 384]$, and the weight enumerator of $\mathcal{C}_{\mathcal{F}}$ is
\begin{eqnarray*}
W_{\mathcal{C}_{\mathcal{F}}}(z)&=&1+2Z^{384}+2Z^{576}+2Z^{720}+2Z^{816}
 +512Z^{4311}+32Z^{4329}+1408Z^{4338}+896Z^{4344}+512Z^{4347}\\
 &+&5824Z^{4356}+3584Z^{4359}+896Z^{4362}+7168Z^{4365}+2240Z^{4368}
 +3584Z^{4371}+12896Z^{4374}\\
 &+&6528Z^{4383}+2048Z^{4386}
 +3584Z^{4389}+5824Z^{4392}+512Z^{4395}+32Z^{4455}
 +224Z^{4473}+224Z^{4485}\\
 &+&224Z^{4500}+32Z^{4512}
 +224Z^{4527}+32Z^{4653}.
\end{eqnarray*}
Clearly, $\frac{w_{min}}{w_{max}}=\frac{384}{4653}<\frac{2}{3}$. The parameters of the code $\mathcal{C}_{\mathcal{F}}^{\bot}$ in Theorem \ref{thm11} is $[6560,6550,2]$.
\end{exmp}
As can be seen from the above example, $\mathcal{E}_h$ is extended by adding zero coordinates to obtain $\mathcal{E}$ satisfying the conditions of Theorem \ref{thm10}, the positions at which these zero coordinates are added will affect the parameters and weight enumerator of $\mathcal{C}_{\mathcal{F}}$. Therefore, in general, it is difficult to determine the weight enumerator of $\mathcal{C}_{\mathcal{F}}$ in Theorem \ref{thm9}.

\section{Conclusions}

In this paper, we first develop a general construction of ternary linear codes of dimension $n+r$, and derive a necessary and sufficient
condition for the resulting codes to be minimal. We next use index codes based on partial spreads to obtain an explicit family of these codes,
and derive the minimality condition for this family from the perspective of the index code. With three classes of index codes satisfying this
condition, we then obtain the parameters and weight enumerators of the corresponding minimal codes, which violate the Ashikhmin-Barg condition, and determine the access structure types of the secret sharing schemes based on their dual codes. We finally use index codes based on
Krawtchouk polynomials to construct the second explicit family of these codes, and establish the index-code condition that guarantees its
minimality. With a class of index codes satisfying this condition, we obtain the parameters of the corresponding minimal code violating the
Ashikhmin-Barg condition, and determine the access structure type of the secret sharing scheme based on its dual code.

Based on the generic construction (\ref{shizi2}), it would be interesting to find more minimal ternary linear codes violating the Ashikhmin-Barg condition. Another interesting research challenge is to extend the construction (\ref{shizi2}) to $\mathbb{F}_p$ to obtain minimal linear codes violating the Ashikhmin-Barg condition, where $p$ is an odd prime.

\end{document}